\documentclass[nopreprint,onecolumn,showkeys]{jasatex}

\usepackage[english]{babel}
\usepackage[utf8x]{inputenc}
\usepackage[T1]{fontenc}
\usepackage{graphicx}
\usepackage{listings}
\usepackage[top=2cm, bottom=2cm, left=1.5cm , right=1.5cm]{geometry}
\usepackage{amsmath}
\usepackage{amsfonts}
\usepackage{cases}
\usepackage{amssymb}
\usepackage{bm}
\newtheorem{proposition}{Proposition}
\begin{document}

\title[Biot-JKD poroelastic waves in 2D heterogeneous media]{A time-domain numerical method 
for Biot-JKD poroelastic waves \\
in 2D heterogeneous media}

\author{Emilie Blanc}\email[Corresponding author: ]{eblanc@lma.cnrs-mrs.fr}
\author{Guillaume Chiavassa}\email[Corresponding author: ]{gchiavassa@centrale-marseille.fr}
\author{Bruno Lombard}\email[Corresponding author: ]{lombard@lma.cnrs-mrs.fr}
\affiliation{Laboratoire de M\'ecanique et d'Acoustique, UPR 7051, CNRS, Marseille\\
Laboratoire M2P2, UMR 7340, Aix-Marseille Universit\'e -- \'Ecole Centrale Marseille}

\date{v0.1 - \today}

\begin{abstract}
An explicit finite-difference scheme is presented for solving the two-dimensional Biot equations of poroelasticity across the full range of frequencies. The key difficulty is to discretize the Johnson-Koplik-Dashen (JKD) model which describes the viscous dissipations in the pores. Indeed, the time-domain version of Biot-JKD model involves order $1/2$ shifted fractional derivatives which amounts to a time convolution product. To avoid storing the past values of the solution, a diffusive representation of fractional derivatives is used: the convolution kernel is replaced by a finite number of memory variables that satisfy local-in-time ordinary differential equations. The coefficients of the diffusive representation follow from an optimization procedure of the dispersion relation. Then, various methods of scientific computing are applied: the propagative part of the equations is discretized using a fourth-order ADER scheme, whereas the diffusive part is solved exactly. An immersed interface method is implemented to discretize the geometry on a Cartesian grid, and also to enforce the jump conditions at interfaces. Numerical experiments are proposed in various realistic configurations.
\end{abstract}



\maketitle

\section{Introduction}\label{sec:intro}

Two frequency regimes have to be distinguished when dealing with poroelastic waves. In the low-frequency range (LF),\cite{BIOT56-A} the viscous efforts are proportional to the relative velocity of the fluid and the solid matrix. In the high-frequency range (HF), modeling the dissipation is a more delicate task: Biot first presented an expression for particular pore geometries.\cite{BIOT56-B} In 1987, Johnson-Koplik-Dashen (JKD) \cite{JKD87} published a general expression for the dissipation in the case of random pores. The viscous efforts depend in this model on the square root of the frequency. In the time domain, time fractional derivatives are then introduced, which involves convolution products with singular kernels. Transient analytical solutions of Biot-JKD model have been derived in academic geometries and homogeneous media.\cite{FELLAH04} When dealing with more complex geometries, numerical methods are required. 

Many time-domain simulation methods  have been developed in the LF regime: see Ref. ~\onlinecite{CARCIONE10} and the introduction to Ref. ~\onlinecite{CHIAVASSA11} for general reviews. In the HF regime, the fractional derivatives greatly complicate the numerical modeling of the Biot-JKD equations. The past values of the solution are indeed required in order to evaluate the convolution products, which means that the time evolution of the solution must be stored. It greatly increases the memory requirements and makes large-scale simulations impossible. To our knowledge, two approaches have been proposed so far in the literature. The first approach consists in discretizing the convolution products,\cite{MASSON10} and the second one is based on the use of a diffusive representation of the fractional derivative.\cite{HANYGA05} In the latter approach, the convolution product is replaced by a continuum of diffusive variables - or memory variables - satisfying local differential equations.\cite{MATIGNON10} This continuum is then discretized using appropriate quadrature formulas, resulting in the Biot-DA (diffusive approximation) model. 

In Ref. ~\onlinecite{BLANC12}, the diffusive approach was followed in a one-dimensional homogeneous configuration. Compared with Ref. ~\onlinecite{HANYGA05}, important improvements were introduced: good representation of the viscous dissipation on the whole range of frequencies; optimization of the model on the range of interest; estimation of the computational effort in terms of the required accuracy. The goal of the present work is to extend the algorithm to two-dimensional heterogeneous porous media with interfaces.

This paper is organized as follows. The original Biot-JKD model is briefly outlined in section \ref{sec:phys} and the principles underlying the diffusive representation of fractional derivatives are described. The decrease of energy and the dispersion analysis are addressed. In section \ref{sec:DA}, the method used to discretize the diffusive model is presented: the diffusive approximation thus obtained is easily treatable by computers. Following a similar approach than in viscoelasticity,\cite{GROBY11} the coefficients of the model are determined using an optimization procedure in the frequency range of interest, giving an optimum number of additional computational arrays. The numerical modeling is addressed in section \ref{sec:num}, where the equations of evolution are split into two parts: a propagative part, which is discretized using a fourth-order scheme for hyperbolic equations, and a diffusive part, which is solved exactly. An immersed interface method accounts for the jump conditions and for the geometry of the interfaces, preventing from the usual limitations of finite differences on a Cartesian grid. Numerical experiments performed with realistic values of the physical parameters are presented in section \ref{sec:exp}. In section \ref{sec:conclu}, a conclusion is drawn and some futures lines of research are given. 
 
%------------------------------------------------------------------------------------------
%------------------------------------------------------------------------------------------

\section{Physical modeling}\label{sec:phys}

\subsection{Biot model}\label{sec:phys:Biot}

The Biot model describes the propagation of mechanical waves in a macroscopic porous medium consisting of a solid matrix saturated with a fluid circulating freely through the pores.\cite{BIOT56-A,BOURBIE87,CARCIONE07} It is assumed that
\begin{itemize}
\item the wavelengths are large in comparison with the diameter of the pores;
\item the amplitude of the perturbations is small;
\item the elastic and isotropic matrix is completely saturated with a single fluid phase;
\item the thermo-mechanical effects are neglected.
\end{itemize}
This model involves 10 physical parameters: the density $\rho_f$ and the dynamic viscosity $\eta$ of the fluid; the density $\rho_s$ and the shear modulus $\mu$ of the elastic skeleton; the porosity $0<\phi<1$, the tortuosity $a\geq 1$, the absolute permeability $\kappa$, the Lam\'e coefficient $\lambda_f$ and the two Biot's coefficients $\beta$ and $m$ of the saturated matrix. The following notations are introduced
\begin{equation}
\begin{array}{l}
\displaystyle
\rho_w = \frac{a}{\phi}\,\rho_f\mbox{,}\quad \rho = \phi\,\rho_f + (1-\phi)\,\rho_s\mbox{,}\quad \chi = \rho\,\rho_w - \rho_f^2 >0,\quad\lambda_0 = \lambda_f-m\,\beta^2.
\end{array}
\label{eq:parametre_biot}
\end{equation}
Taking $\bf u_s$ and $\bf u_f$ to denote the solid and fluid displacements, the unknowns in 2D are the elastic velocity ${\bf v_s}=\frac{\partial\,\bf u_s}{\partial\,t}$, the filtration velocity ${\bf w}=\frac{\partial\,{\bf{\cal W}}}{\partial\,t}=\phi\,\frac{\partial }{\partial\,t}\,({\bf u_f-u_s})$, the elastic stress tensor $\bm{\sigma}$, and the acoustic pressure $p$. The constitutive laws are
\begin{equation}
\begin{array}{l}
\displaystyle \bm{\sigma} = (\lambda_f\,\mbox{tr}\,\bm{\varepsilon} - \beta\,m\,\xi)\,{\bf I} + 2\,\mu\bm{\varepsilon},\\
[10pt]
\displaystyle p = m\,(-\beta\,\mbox{tr}\,\bm{\varepsilon} + \xi),
\end{array}
\label{eq:biot_comportement}
\end{equation}
where $\bm{\varepsilon} = \frac{1}{2}\,(\nabla{\bf u_s} + \nabla{\bf u_s}^T)$ is the strain tensor and $\xi = -\nabla.{\bf{\cal W}}$ is the rate of fluid change. The symmetry of $\bm{\sigma}$  implies compatibility conditions between spatial derivatives of $\bm{\varepsilon}$, leading to the  Beltrami-Michell equation:\cite{COUSSY95}
\begin{equation}
\begin{array}{l}
\displaystyle \frac{\partial^2\,\sigma_{xy}}{\partial x\,\partial y} = \theta_0\,\frac{\partial^2\,\sigma_{xx}}{\partial x^2} + \theta_1\,\frac{\partial^2\,\sigma_{yy}}{\partial x^2} + \theta_2\,\frac{\partial^2\,p}{\partial x^2} + \theta_1\,\frac{\partial^2\,\sigma_{xx}}{\partial y^2} + \theta_0\,\frac{\partial^2\,\sigma_{yy}}{\partial y^2} + \theta_2\,\frac{\partial^2\,p}{\partial y^2},\\
[10pt]
\displaystyle \theta_0 = -\frac{\lambda_0}{4(\lambda_0 + \mu)},\quad\theta_1 = \frac{\lambda_0 + 2\,\mu}{4(\lambda_0 + \mu)},\quad\theta_2 = \frac{\mu\,\beta}{2(\lambda_0 + \mu)}.
\end{array}
\label{eq:beltrami}
\end{equation}
On the other hand, the conservation of momentum yields
\begin{equation}
\left\lbrace 
\begin{array}{l}
\displaystyle \rho\,\frac{\partial \bf v_s}{\partial t} + \rho_f\,\frac{\partial\bf w}{\partial t} = \nabla \bm{\sigma},\\
[10pt]
\displaystyle \rho_s\,\frac{\partial \bf v_s}{\partial t} + \rho_w\,\frac{\partial \bf w}{\partial t} + \frac{\eta}{\kappa}\,F*{\bf w} = -\nabla p,
\end{array}
\right. 
\label{eq:biot_dynamique}
\end{equation}
where $*$ is the convolution product in time. The second equation of (\ref{eq:biot_dynamique}) is a generalized Darcy law. The quantity $F*{\bf w}$  denotes the viscous dissipation induced by the relative motion between the fluid and the elastic skeleton.

%------------------------------------------------------------------------------------------

\subsection{High frequency dissipation: the JKD model}\label{sec:phys:JKD}

The frontier between the low-frequency range (LF) and the high-frequency range (HF) is reached when the viscous efforts are similar to the inertial effects. The transition frequency is given by \cite{BIOT56-A,BOURBIE87} 
\begin{equation}
f_c = \frac{\eta\,\phi}{2\,\pi\,a\,\kappa\,\rho_f}=\frac{\textstyle \omega_c}{\textstyle 2\,\pi}.
\label{eq:fc}
\end{equation}
In LF, the flow in the pores is of the Poiseuille type, and dissipation efforts in the second equation of (\ref{eq:biot_comportement}) are given by
\begin{equation}
F(t)= \delta(t) \Longleftrightarrow F(t)*{\bf w}(x,y,\,t)={\bf w}(x,y,\,t),
\end{equation}
where $\delta$ is the Dirac distribution. In HF, the width of the viscous boundary-layer is small in comparison with the size of the pores, and modeling the dissipation process is a more complex task. Here we adopt the widely-used model proposed by Johnson-Koplik-Dashen (JKD) in 1987, which is valid for random networks of pores with constant radii.\cite{JKD87} The only additional parameter is the viscous characteristic length $\Lambda$. We take
\begin{equation}
P=\frac{4\,a\,\kappa}{\phi\,\Lambda^2}\mbox{,}\qquad\Omega = \frac{\omega_c}{P} = \frac{\eta\,\phi^2\,\Lambda^2}{4\,a^2\,\kappa^2\,\rho_f},
\label{eq:coef_hf}
\end{equation}
where $P$ is the Pride number (typically $P \approx 1/2$). Based on the Fourier transform in time, $\widehat{F}(\omega) = \int _{\mathbb R} F(t)e^{-i\omega t}\,dt$, the frequency correction given by the JKD model can be written
\begin{equation}
\begin{array}{ll}
\widehat{F}(\omega) & \displaystyle = \left( 1+i\,\omega\,\frac{4\,a^2\,\kappa^2\,\rho_f}{\eta\,\Lambda^2\,\phi^2}\right) ^{1/2},\\
[10pt]
& \displaystyle = \left( 1+i\,P\,\frac{\omega}{\omega_c}\right) ^{1/2},\\
[10pt]
& \displaystyle = \frac{1}{\sqrt{\Omega}}\,(\Omega +i\,\omega)^{1/2}.
\end{array}
\label{eq:F_omega}
\end{equation}
This correction is the simplest function satisfying the LF and HF limits of the dynamic permeability.\cite{JKD87} Therefore, the term $F(t)*{\bf w}(x,y,t)$ involved in the second equation of (\ref{eq:biot_comportement}) is
\begin{equation}
\begin{array}{ll}
F(t)*{\bf w}(x,y,t) & \displaystyle = \mathcal{F}^{-1}\left( \frac{1}{\sqrt{\Omega}}\,(\Omega + i\,\omega)^{1/2}\widehat{\bf w}(x,y,\omega)\right),\\
[10pt]
& \displaystyle = \frac{1}{\sqrt{\Omega}}\,(D+\Omega)^{1/2}{\bf w}(x,y,t).
\end{array}
\label{eq:F_t}
\end{equation}
The operator $D^{1/2}$ is a shifted order 1/2 time fractional derivative, generalizing the usual derivative characterized by $\frac{\partial\,\bf w}{\partial\,t} = \mathcal{F}^{-1}\left( i\,\omega\,\widehat{\bf w}(x,y,\omega)\right)$. The notation $(D+\Omega)^{1/2}$ accounts for the shift $\Omega$ in (\ref{eq:F_t}).

%------------------------------------------------------------------------------------------

\subsection{The Biot-JKD equations of evolution}\label{sec:phys:EDP}

Based on (\ref{eq:biot_comportement}), (\ref{eq:biot_dynamique}) and (\ref{eq:F_t}), the Biot-JKD equations can be written
\begin{equation}
\left\lbrace 
\begin{array}{l}
\displaystyle \rho\,\frac{\partial\,\bf v_s}{\partial\,t} + \rho_f\,\frac{\partial\,\bf w}{\partial\,t}=\nabla\bm{\sigma},\\
[10pt]
\displaystyle \rho_f\,\frac{\partial\,\bf v_s}{\partial\,t} + \rho_w\,\frac{\partial\,\bf w}{\partial\,t} + \frac{\eta}{\kappa}\,\frac{1}{\sqrt{\Omega}}\,(D+\Omega)^{1/2}\,{\bf w} = -\nabla p,\\
[10pt]
\displaystyle \bm{\sigma} = (\lambda_f\,\mbox{tr}\,\bm{\varepsilon} - \beta\,m\,\xi){\bf I} + 2\,\mu\,\bm{\varepsilon},\\
[10pt]
\displaystyle p = m\,(-\beta\,\mbox{tr}\,\bm{\varepsilon} + \xi).
\end{array}
\right. 
\label{eq:LCBiot}
\end{equation}
We rearrange this system by separating  
$\frac{\partial\,\bf v_s}{\partial\,t}$ and $\frac{\partial\,\bf w}{\partial\,t}$ in the first two equations of (\ref{eq:LCBiot}) and by using the definitions of $\bm{\varepsilon}$ and $\xi$. Taking
\begin{equation}
\gamma = \frac{\eta}{\kappa}\,\frac{\rho}{\chi}\,\frac{1}{\sqrt{\Omega}},
\label{eq:coef_gamma_omega_biot}
\end{equation}
one obtains the following system of equations of evolution 
\begin{equation}
\left\lbrace 
\begin{array}{l}
\displaystyle \frac{\partial\,v_{sx}}{\partial\,t} - \frac{\rho _w}{\chi}\,\left( \frac{\partial\,\sigma_{xx}}{\partial\,x}+\frac{\partial\,\sigma_{xy}}{\partial\,y}\right)  - \frac{\rho _f}{\chi}\,\frac{\partial\,p}{\partial\,x} = \frac{\rho _f}{\rho}\,\gamma\,(D+\Omega)^{1/2}w_{x}+f_{v_{sx}},\\
[15pt]
\displaystyle \frac{\partial\,v_{sy}}{\partial\,t} - \frac{\rho _w}{\chi}\,\left( \frac{\partial\,\sigma_{xy}}{\partial\,x}+\frac{\partial\,\sigma_{yy}}{\partial\,y}\right)  - \frac{\rho _f}{\chi}\,\frac{\partial\,p}{\partial\,y} = \frac{\rho _f}{\rho}\,\gamma\,(D+\Omega)^{1/2}w_{y}+f_{v_{sy}},\\
[15pt]
\displaystyle \frac{\partial\,w_x}{\partial\,t} + \frac{\rho _f}{\chi}\,\left( \frac{\partial\,\sigma_{xx}}{\partial\,x}+\frac{\partial\,\sigma_{xy}}{\partial\,y}\right)  + \frac{\rho }{\chi}\,\frac{\partial\,p}{\partial\,x} = -\,\gamma\,(D+\Omega)^{1/2}w_x+f_{w_x},\\
[15pt]
\displaystyle \frac{\partial\,w_y}{\partial\,t} + \frac{\rho _f}{\chi}\,\left( \frac{\partial\,\sigma_{xy}}{\partial\,x}+\frac{\partial\,\sigma_{yy}}{\partial\,y}\right)  + \frac{\rho }{\chi}\,\frac{\partial\,p}{\partial\,y} = -\,\gamma\,(D+\Omega)^{1/2}w_y+f_{w_y},\\
[15pt]
\displaystyle \frac{\partial\,\sigma_{xx}}{\partial\,t} - (\lambda_f + 2\mu)\,\frac{\partial\,v_{sx}}{\partial\,x} - m\,\beta \,\frac{\partial\,w_x}{\partial\,x}-\lambda_f\,\frac{\partial\,v_{sy}}{\partial\,y}-m\,\beta\,\frac{\partial\,w_y}{\partial\,y}=f_{\sigma_{xx}},\\
[15pt]
\displaystyle \frac{\partial\,\sigma_{xy}}{\partial\,t}-\mu\,\left( \frac{\partial\,v_{sy}}{\partial\,x}+\frac{\partial\,v_{sx}}{\partial\,y}\right) =f_{\sigma_{xy}},\\
[15pt]
\displaystyle \frac{\partial\,\sigma_{yy}}{\partial\,t}-\lambda_f\,\frac{\partial\,v_{sx}}{\partial\,x}-m\,\beta\,\frac{\partial\,w_x}{\partial\,x}- (\lambda_f + 2\mu)\,\frac{\partial\,v_{sy}}{\partial\,y} - m\,\beta \,\frac{\partial\,w_y}{\partial\,y}=f_{\sigma_{yy}},\\
[15pt]
\displaystyle \frac{\partial\,p}{\partial\,t} + m\, \left(  \beta\,\frac{\partial\,v_{sx}}{\partial\,x} +\frac{\partial\,w_x}{\partial\,x}+\beta\,\frac{\partial\,v_{sy}}{\partial\,y} +\frac{\partial\,w_y}{\partial\,y} \right)  =f_p.
\end{array}
\right. 
\label{eq:S1}
\end{equation}
Terms $f_{v_{sx}}$, $f_{v_{sy}}$, $f_{w_x}$, $f_{w_y}$, $f_{\sigma_{xx}}$, $f_{\sigma_{xy}}$, $f_{\sigma_{yy}}$ and $f_p$ have been  to simulate sources.
%where $g(t)=\frac{d G}{dt}(t)$.
%where we have introduced a source term $s(x,\,t)=G(t)\,h(x)$ in the stress constitutive law. 

%------------------------------------------------------------------------------------------

\subsection{The diffusive representation}\label{sec:phys:DR}

Taking
\begin{equation}
D_{\Omega}{\bf w}(x,y,t) = \frac{\partial \,{\bf w}}{\partial \,t}+\Omega\,{\bf w},
\end{equation}
the shifted fractional derivative (\ref{eq:F_t}) can be written \cite{DUBOIS10}
\begin{equation}
(D+\Omega )^{1/2}{\bf w}(x,y,t) = \frac{1}{\sqrt{\pi}}\,\int _0 ^t \frac{e^{-\Omega (t-\tau)}}{\sqrt{t-\tau}}D_{\Omega}{\bf w}(x,y,\tau)\,d\tau.
\label{eq:Dfrac}
\end{equation}
The operator $(D+\Omega)^{1/2}$  is not local in time and involves the entire time history of ${\bf w}$. As we will see in section \ref{sec:DA}, a different way of writing this derivative is more convenient for numerical evaluation. Based on Euler's $\Gamma$ function, the diffusive representation of the totally monotone function $\frac{1}{\sqrt{t}}$ is\cite{MATIGNON10}
\begin{equation}
\displaystyle \frac{1}{\sqrt{t}} = \frac{1}{\sqrt{\pi}} \int _0 ^\infty\,\frac{1}{\sqrt{\theta}}\,e^{-\theta t}d\theta .
\label{eq:fonction_diffu}
\end{equation}
Substituting (\ref{eq:fonction_diffu}) into (\ref{eq:Dfrac}) gives
\begin{equation}
\begin{array}{ll}
\displaystyle (D+\Omega )^{1/2}{\bf w}(x,y,t) & \displaystyle = \frac{1}{\pi}\,\int _0 ^t \int _0 ^{\infty}\frac{1}{\sqrt{\theta}}\,e^{-\theta (t-\tau)}\,e^{-\Omega (t-\tau)}\,D_{\Omega}{\bf w}(x,y,\tau)\,d\theta\, d\tau,\\
[20pt]
& \displaystyle = \frac{1}{\pi}\, \int _0 ^{\infty}\frac{1}{\sqrt{\theta}}\,\bm{\psi}(x,y,\theta,t)\,d\theta,
\end{array}
\label{eq:derivee_frac}
\end{equation}
where the diffusive variable is defined as
\begin{equation}
\bm{\psi}(x,y,\theta,t)=\int_0^t e^{-(\theta + \Omega )(t-\tau)}\,D_{\Omega}{\bf w}(x,y,\tau )\,d\tau.
\label{eq:variable_diffu}
\end{equation}
For the sake of clarity, the dependence on $\Omega$ and ${\bf w}$ is omitted in $\bm{\psi}$. From (\ref{eq:variable_diffu}), it follows that the diffusive variable $\bm{\psi}$ satisfies the ordinary differential equation
\begin{equation}
\left\lbrace 
\begin{array}{l}
\displaystyle \frac{\partial\,\bm{\psi}}{\partial\,t} = -(\theta + \Omega )\,\bm{\psi} + D_{\Omega}{\bf w}, \\
[10pt]
\displaystyle \bm{\psi}(x,y,\theta,0) = 0.
\end{array}
\right.
\label{eq:EDO_psi}
\end{equation}
The diffusive representation therefore transforms a non-local problem (\ref{eq:Dfrac}) into a continuum of local problems (\ref{eq:EDO_psi}). It should be emphasized at this point that no approximation have been made up to now. The computational advantages of the diffusive representation will be seen in sections \ref{sec:DA} and \ref{sec:exp}, where the discretization of (\ref{eq:derivee_frac}) and (\ref{eq:EDO_psi}) will yield a numerically tractable formulation.

%------------------------------------------------------------------------------------------

\subsection{Energy of Biot-JKD}\label{sec:phys:NRJ}

Now, we express the energy of the Biot-JKD model (\ref{eq:LCBiot}). This result generalizes the analysis performed in the 1D case in the Ref. ~\onlinecite{BLANC12}.

\begin{proposition}
Setting $\bm{C}$ the $4 \times 4$ poroelastic tensor such that $\bm{\sigma} = \bm{C}\,\bm{\varepsilon} - \beta\,p\,\bm{I}$ in (\ref{eq:biot_comportement}), let
$$
E=E_1+E_2+E_3,
$$
with
\begin{equation}
\begin{array}{lll}
E_1  & = & \displaystyle \frac{\textstyle 1}{\textstyle 2}\,\int_{\mathbb{R}^2}\left(\rho\,{\bf v}_s^2+\rho_w\,{\bf w}^2+2\,\rho_f\,{\bf v}_s.{\bf w}\right) dx\,dy,\\ 
[20pt]
E_2 & = & \displaystyle \frac{\textstyle 1}{\textstyle 2}\,\int_{\mathbb{R}^2}\left( {\bf C}\,\bm{\varepsilon}\,:\,\bm{\varepsilon} + \frac{\textstyle 1}{\textstyle m}\,p^2\right)\,dx\,dy,\\
[20pt]
E_3 & = & \displaystyle \frac{\textstyle 1}{\textstyle 2}\,\int_{\mathbb{R}^2}\int_{\theta\in\mathbb{R}^+}\frac{\eta}{\kappa}\,\frac{1}{\pi}\,\frac{1}{\sqrt{\Omega\,\theta}}\,\frac{1}{\theta+2\,\Omega}\,({\bf w}-\bm{\psi})^2\,d\theta\,dx\,dy.
\end{array}
\label{eq:energie}
\end{equation}
Without any source terms, $E$ is an energy which satisfies
\begin{equation}
\frac{\textstyle dE}{\textstyle dt} = -\int_{\mathbb{R}^2}\int_{\theta\in\mathbb{R}^+}\frac{\eta}{\kappa}\,\frac{1}{\pi}\,\frac{1}{\sqrt{\Omega\,\theta}}\,\frac{1}{\theta+2\,\Omega}\,\left( \Omega\,{\bf w}^2+(\theta+\Omega)\,\bm{\psi}^2\right) d\theta\,dx\,dy \;\leq\; 0.
\label{eq:dEdt}
\end{equation}
\label{prop:nrj}
\end{proposition}

Proposition \ref{prop:nrj} calls for the following comments:
\begin{itemize}
\item the Biot-JKD model is well-posed;
\item when the viscosity of the saturating fluid is neglected ($\eta=0$), the energy of the system is conserved;
\item the terms in (\ref{eq:energie}) have a clearly physical significance: $E_1$ is the kinetic energy, and $E_2$ is the potential energy. The term $E_3$ corresponds to the kinetic energy resulting from the relative motion of the fluid inside the pores.
\end{itemize}

%-----------------------------------------------------------------

\subsection{Dispersion analysis}\label{sec:phys:dispersion}

\begin{figure}[htbp]
\begin{center}
\begin{tabular}{cc}
%phase velocity of the fast compressional wave & attenuation of the fast compressional wave\\
phase velocity of the & attenuation of the\\
fast compressional wave & fast compressional wave\\
\includegraphics[scale=0.34]{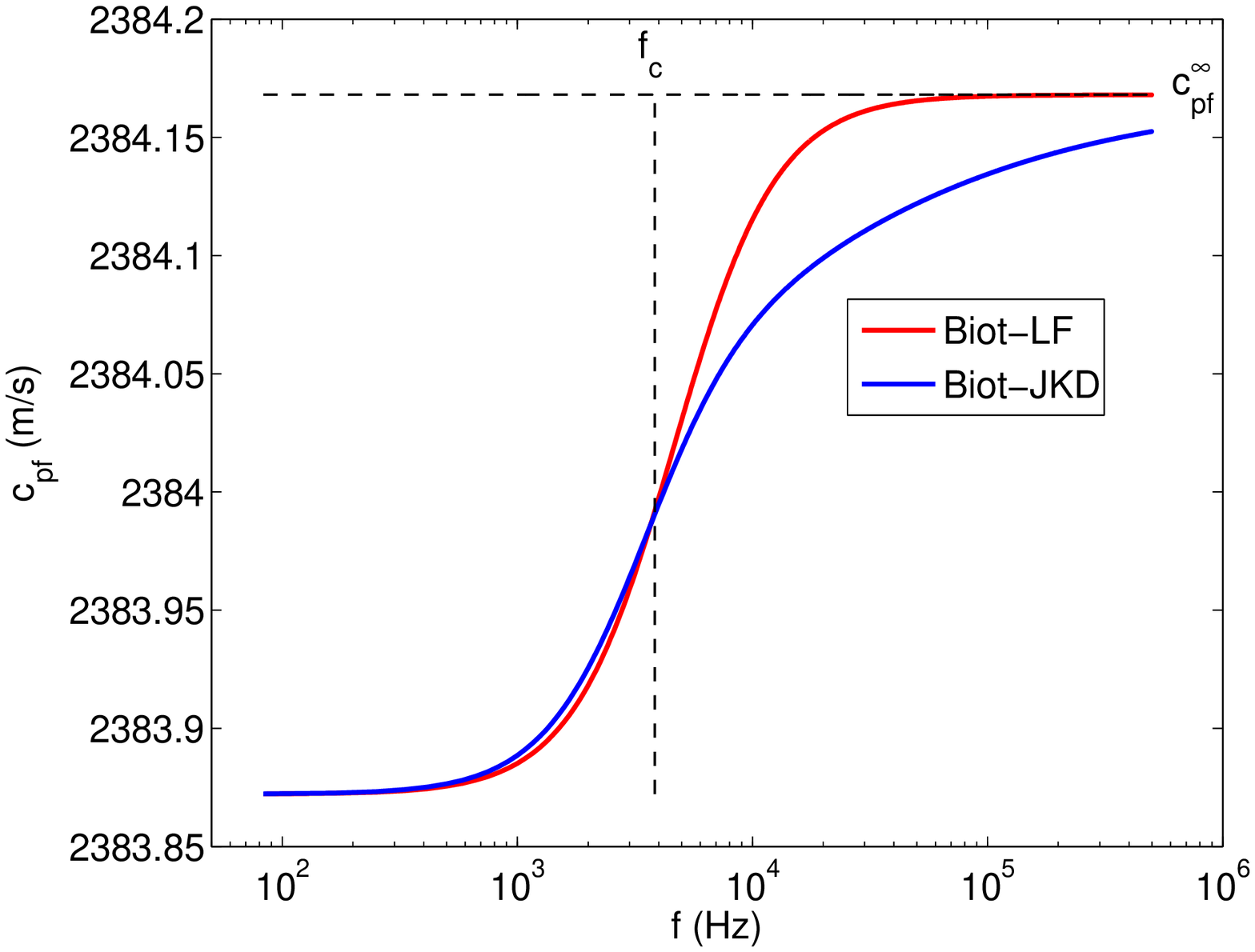} &
\includegraphics[scale=0.34]{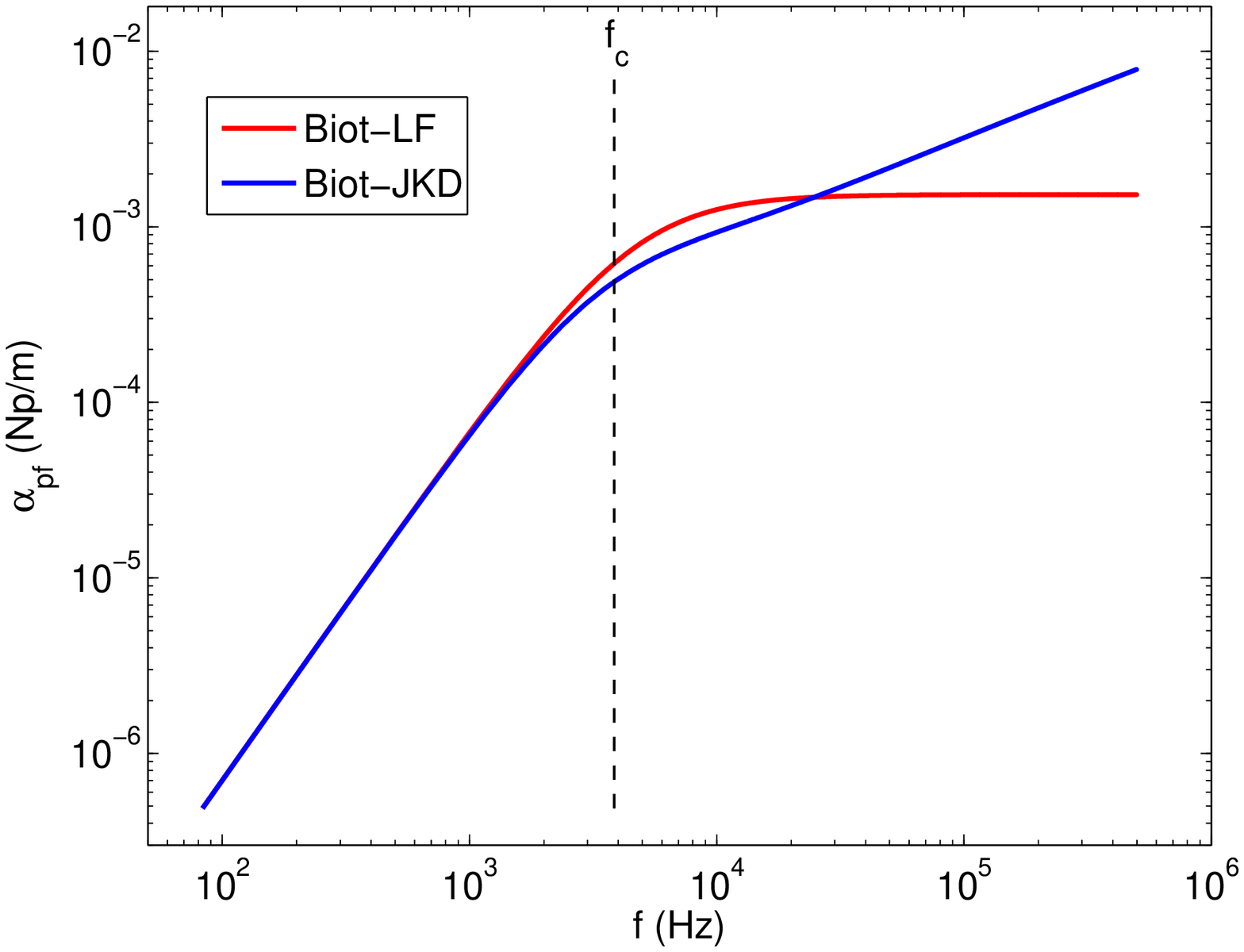}\\
phase velocity of the shear wave & attenuation of the shear wave\\
\includegraphics[scale=0.34]{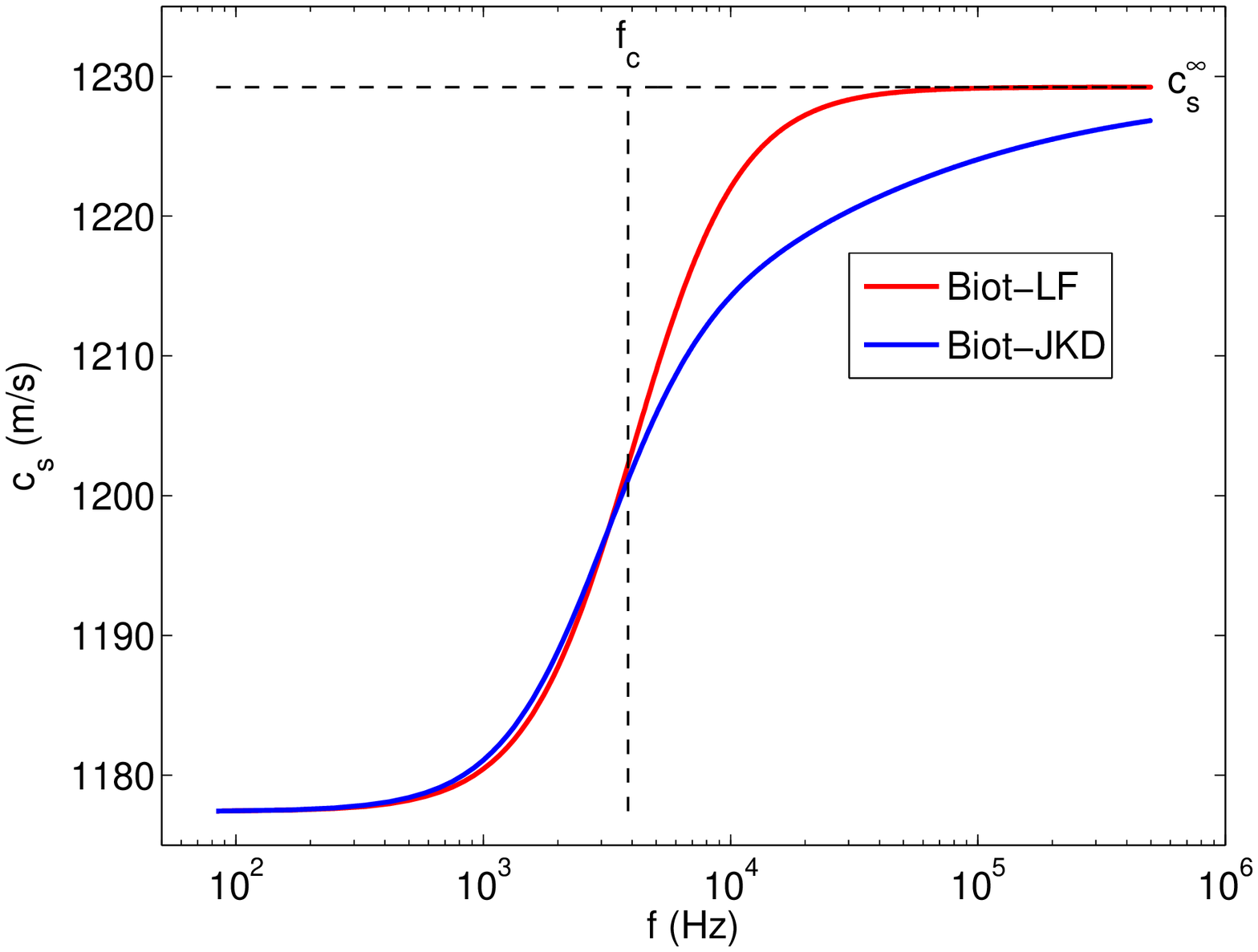} &
\includegraphics[scale=0.34]{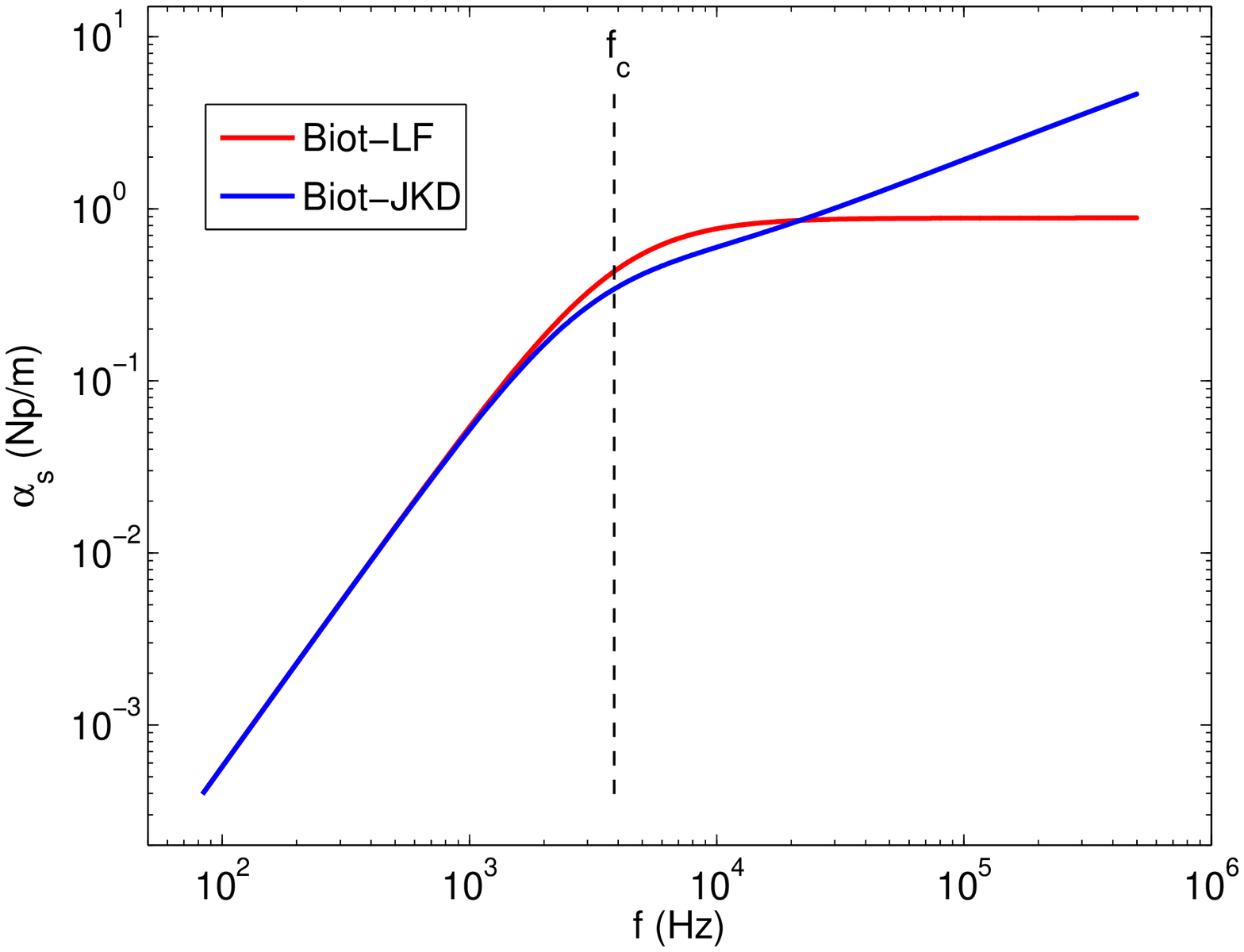}\\
phase velocity of the & attenuation of the\\
slow compressional wave & slow compressional wave\\
\includegraphics[scale=0.34]{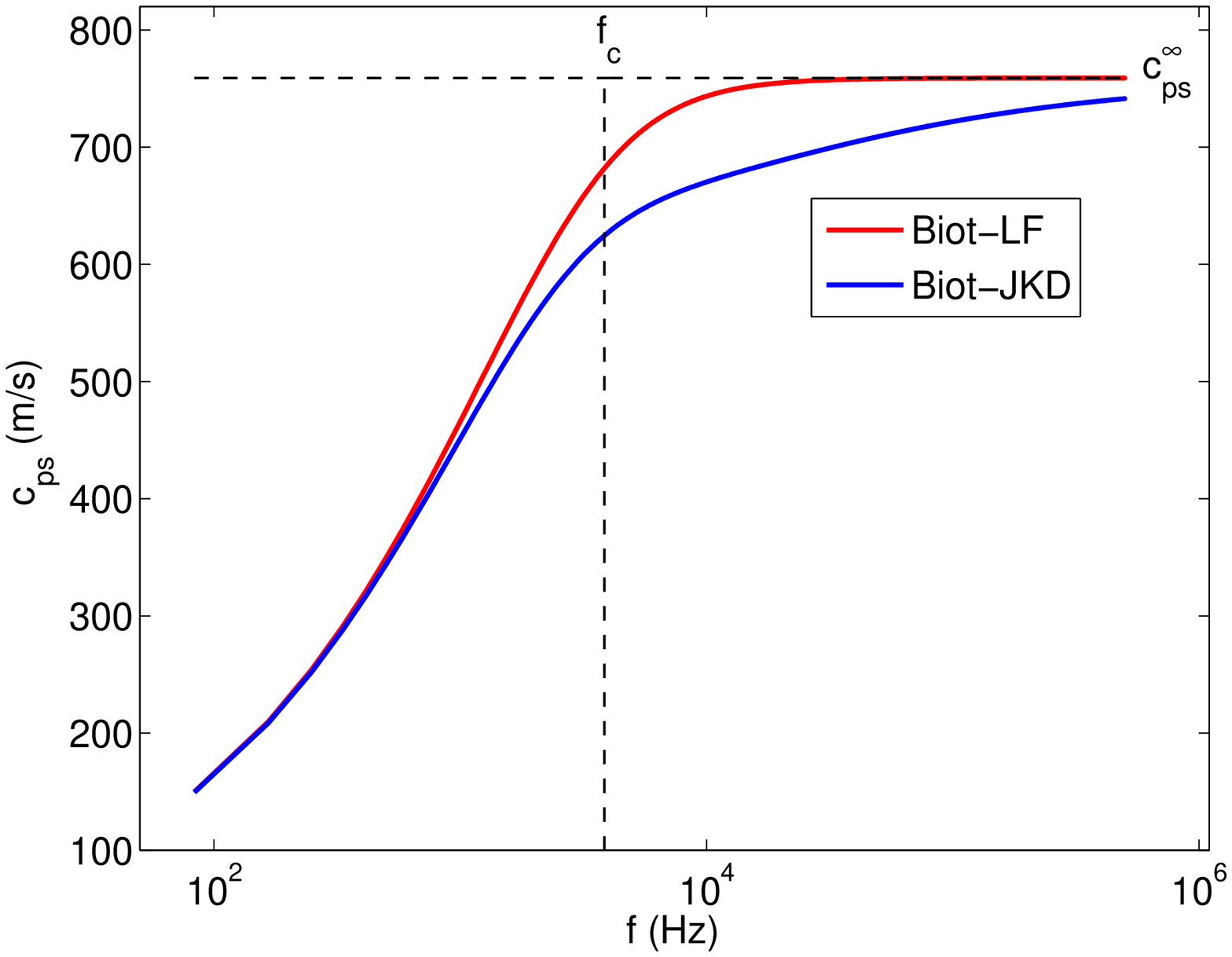} &
\includegraphics[scale=0.34]{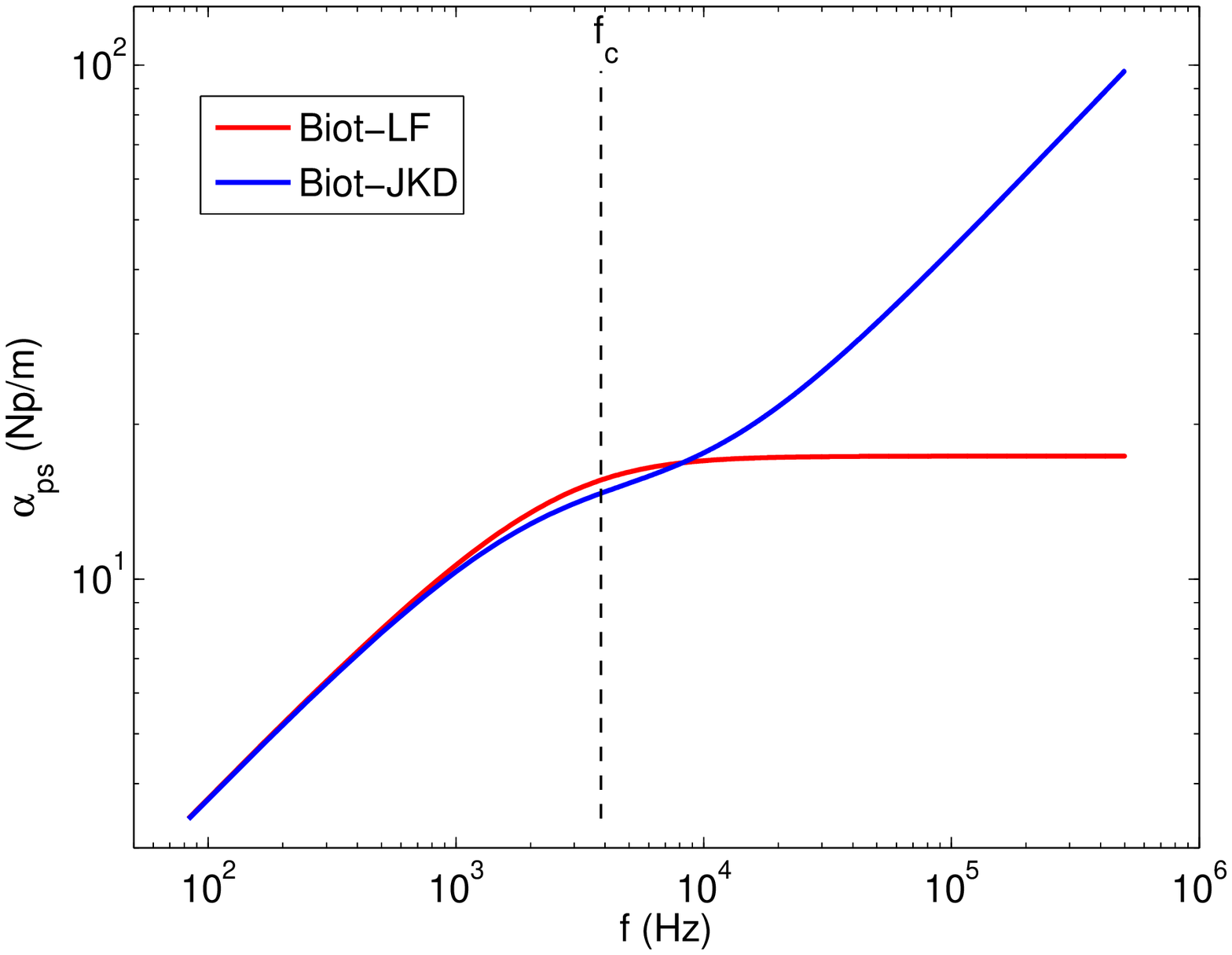}\\
\end{tabular}
\end{center}
\caption{Dispersion curves: comparison between Biot-LF and Biot-JKD.}
\label{fig:dispersion_bf}
\end{figure}

A plane wave ${\bf d}e^{i(\omega t-{\bf k}.{\bf r})}$ is injected in (\ref{eq:S1}), where ${\bf k} = k\,{\bf e}$ and $\bf d$ are the wavevector and the polarization, respectively; $\bf r$ is the position, $\omega = 2\,\pi\,f$ is the angular frequency and $f$ is the frequency. If $\bf d$ is collinear with $\bf k$, the dispersion relation of compressional waves is obtained. Setting
\begin{equation}
\left\lbrace \begin{array}{ll}
\displaystyle D_4 & = m\,(\lambda_0+2\,\mu),\\
[12pt]
\displaystyle D_2(\omega) & = \displaystyle -\left( (\lambda_f+2\,\mu)\,\rho_w + m\,(\rho-2\,\rho_f\,\beta)\right) \,\omega^2 + i\,\omega\,\frac{\eta}{\kappa}\,\widehat{F}(\omega)\,(\lambda_f+2\,\mu),\\
[10pt]
\displaystyle D_0(\omega) & = \displaystyle \chi\,\omega^4 - i\,\omega^3\,\frac{\eta}{\kappa}\,\rho\,\widehat{F}(\omega),
\end{array}
\right. 
\label{eq:coef_relation_dis_P}
\end{equation}
the dispersion relation of compressional waves takes the form
\begin{equation}
D_e(k,\omega ) = D_4\,k^4 + D_2(\omega)\,k^2 + D_0(\omega) = 0.
\label{eq:relation_dispersion_P}
\end{equation}
If $\bf d$ is orthogonal with $\bf k$, the dispersion relation of the shear wave is obtained. Setting
\begin{equation}
\left\lbrace \begin{array}{ll}
\displaystyle D_4(\omega) & \displaystyle = \omega^2\,(\rho+\phi\,\rho_f\,(a-2))-i\,\omega\,\phi^2\,\frac{\eta}{\kappa}\,\widehat{F}(\omega),\\
[12pt]
\displaystyle D_2(\omega) & \displaystyle = -\omega^2\,\phi\,\rho_f\,(a-1) + i\,\omega\,\phi^2\,\frac{\eta}{\kappa}\,\widehat{F}(\omega),\\
[10pt]
\displaystyle D_0(\omega) & \displaystyle = \omega^2\,\phi\,\rho_f\,a - i\,\omega\,\phi^2\,\frac{\eta}{\kappa}\,\widehat{F}(\omega),
\end{array}
\right. 
\label{eq:coef_relation_dis_shear}
\end{equation}
the dispersion relation of shear wave takes the form
\begin{equation}
k = \frac{1}{\sqrt{\mu}}\,\left( \frac{D_4\,D_0-D_2^2}{D_0}\right)^{1/2}.
\label{eq:relation_dispersion_shear}
\end{equation}
Expressions (\ref{eq:coef_relation_dis_P})-(\ref{eq:relation_dispersion_P})-(\ref{eq:coef_relation_dis_shear})-(\ref{eq:relation_dispersion_shear}) are valid in the case of both the Biot-LF and Biot-JKD  models with the frequency correction defined by
\begin{equation}
\widehat{F}(\omega ) = 
\left\lbrace 
\begin{array}{lll}
\displaystyle \widehat{F}_{LF}(\omega) & \displaystyle = 1 & \mbox{Biot-LF},\\
[10pt]
\displaystyle \widehat{F}_{JKD}(\omega) & \displaystyle = \frac{1}{\sqrt{\Omega}}\,(\Omega +i\,\omega)^{1/2} & \mbox{Biot-JKD}.
\end{array}
\right. 
\label{eq:fonction_correction}
\end{equation}
The solutions $k_{pf}$, $k_{ps}$ of (\ref{eq:relation_dispersion_P}) and the solution $k_{s}$ of (\ref{eq:relation_dispersion_shear}) give the phase velocities $c_{pf} = \omega/\Re\mbox{e}(k_{pf})$ of the fast compressional wave, $c_{ps} = \omega/\Re\mbox{e}(k_{ps})$ of the slow compressional wave, and $c_{s} = \omega/\Re\mbox{e}(k_{s})$ of the shear wave, with $0<c_{ps}<c_{pf}$ and $0<c_{s}$. The attenuations $\alpha_{pf} = -\Im\mbox{m}(k_{pf})$, $\alpha_{ps} = -\Im\mbox{m}(k_{ps})$ and $\alpha_{s} = -\Im\mbox{m}(k_{s})$ can also be deduced. Both the phase velocities and the attenuations of Biot-LF and Biot-JKD are strictly increasing functions of the frequency. 
The high frequency limits of phase velocities of compressional waves, $c_{pf}^{\infty}$ and $c_{ps}^{\infty}$,  obtained by diagonalizing the left-hand side of system (\ref{eq:S1}), satisfy the relation
\begin{equation}
\chi\,c^4 - \left( (\lambda_f+2\,\mu)\,\rho_w+m\,(\rho-2\,\rho_f\,\beta) \right)  \,c^2 + m\,(\lambda_0+2\,\mu) = 0,
\label{eq:poly_cara_A}
\end{equation}
and the high frequency limit of phase velocity of the shear wave, $c_{s}^{\infty}$, is
\begin{equation}
c_s^{\infty} = \sqrt{\mu}\,\left( \rho - \frac{\phi\,\rho_f}{a}\right) ^{-1/2}.
\label{eq:poly_cara_A_shear}
\end{equation}
Figure \ref{fig:dispersion_bf} shows the dispersion curves corresponding to the Biot-LF and Biot-JKD models. The physical parameters are those used in the numerical experiments presented in section \ref{sec:exp} (medium $\Omega_0$). Note that the scales are radically different in the case of fast and slow waves. The following properties can be observed:
\begin{itemize}
\item when $f<f_c$, the Biot-JKD and Biot-LF dispersion curves are very similar as might be expected, since $\displaystyle{\lim_{\omega \rightarrow 0} \widehat{F}_{JKD}(\omega)=1}$;
\item the fast compressional wave and the shear wave are almost not affected by the frequency correction $\widehat{F}(\omega)$, while the slow compressional wave is greatly modified;
\item when $f\ll f_c$, the slow compressional wave is almost static. When $f>f_c$, the slow wave propagates but is greatly attenuated.
\end{itemize}

%------------------------------------------------------------------------------------------
%------------------------------------------------------------------------------------------

\section{The Biot-DA (diffusive approximation) model}\label{sec:DA}

The aim of this section is to approximate the Biot-JKD model, using a numerically tractable approach. 

\subsection{The Biot-DA first-order system}\label{sec:DA:EDP}

Using a quadrature formula on $N$ points, with weights $a_{\ell}$ and abscissa $\theta_{\ell}>0$, the diffusive representation (\ref{eq:derivee_frac}) can be approximated by
\begin{equation}
\begin{array}{ll}
\displaystyle (D+\Omega )^{1/2}{\bf w}(x,y,t) & \displaystyle = \frac{1}{\pi}\,\int _0 ^{\infty}\frac{1}{\sqrt{\theta}}\,\bm{\psi} (x,y,t,\theta)\,d\theta, \\
[12pt]
& \displaystyle \simeq \sum \limits _{\ell=1} ^{N}a_{\ell} \,\bm{\psi}(x,y,t,\theta_{\ell}),\\
[12pt]
& \displaystyle \equiv \sum \limits _{\ell=1} ^{N}a_{\ell} \,\bm{\psi}_{\ell} (x,y,t).
\label{eq:derivee_frac_rd}
\end{array}
\end{equation}
>From (\ref{eq:EDO_psi}), the $N$ diffusive variables $\bm{\psi}_{\ell}$ satisfy the ordinary differential equations
\begin{equation}
\left\lbrace 
\begin{array}{l}
\displaystyle \frac{\partial\,\bm{\psi}_{\ell}}{\partial\,t} = -(\theta_{\ell} + \Omega )\,\bm{\psi}_{\ell} + D_{\Omega}{\bf w}, \\
[10pt]
\displaystyle \bm{\psi}_{\ell} (x,y,0) = 0.
\end{array}
\right.
\label{eq:EDO_psi_l}
\end{equation}

The fractional derivatives are replaced by their diffusive approximation (\ref{eq:derivee_frac_rd}) in the JKD model (\ref{eq:S1}). Upon adding the equations (\ref{eq:EDO_psi_l}) and performing some straightforward operations, the Biot-DA system is written as a first-order system in time and space
\begin{equation}
\left\lbrace 
\begin{array}{l}
\displaystyle \frac{\partial\,v_{sx}}{\partial\,t} - \frac{\rho _w}{\chi}\,\left( \frac{\partial\,\sigma_{xx}}{\partial\,x}+\frac{\partial\,\sigma_{xy}}{\partial\,y}\right)  - \frac{\rho _f}{\chi}\,\frac{\partial\,p}{\partial\,x} = \frac{\rho _f}{\rho}\,\gamma\,\sum \limits _{\ell =1}^{N}a_{\ell } \,\psi _{x\ell }+f_{v_{sx}},\\
[15pt]
\displaystyle \frac{\partial\,v_{sy}}{\partial\,t} - \frac{\rho _w}{\chi}\,\left( \frac{\partial\,\sigma_{xy}}{\partial\,x}+\frac{\partial\,\sigma_{yy}}{\partial\,y}\right)  - \frac{\rho _f}{\chi}\,\frac{\partial\,p}{\partial\,y} = \frac{\rho _f}{\rho}\,\gamma\,\sum \limits _{\ell =1}^{N}a_{\ell } \,\psi _{y\ell }+f_{v_{sy}},\\
[15pt]
\displaystyle \frac{\partial\,w_x}{\partial\,t} + \frac{\rho _f}{\chi}\,\left( \frac{\partial\,\sigma_{xx}}{\partial\,x}+\frac{\partial\,\sigma_{xy}}{\partial\,y}\right)  + \frac{\rho }{\chi}\,\frac{\partial\,p}{\partial\,x} = -\gamma\,\sum \limits _{\ell =1}^{N}a_{\ell }\, \psi _{x\ell }+f_{w_x},\\
[15pt]
\displaystyle \frac{\partial\,w_y}{\partial\,t} + \frac{\rho _f}{\chi}\,\left( \frac{\partial\,\sigma_{xy}}{\partial\,x}+\frac{\partial\,\sigma_{yy}}{\partial\,y}\right)  + \frac{\rho }{\chi}\,\frac{\partial\,p}{\partial\,y} = -\gamma\,\sum \limits _{\ell =1}^{N}a_{\ell }\, \psi _{y\ell }+f_{w_y},\\
[15pt]
\displaystyle \frac{\partial\,\sigma_{xx}}{\partial\,t} - (\lambda_f + 2\mu)\,\frac{\partial\,v_{sx}}{\partial\,x} - m\,\beta \,\frac{\partial\,w_x}{\partial\,x}-\lambda_f\,\frac{\partial\,v_{sy}}{\partial\,y}-m\,\beta\,\frac{\partial\,w_y}{\partial\,y}=f_{\sigma_{xx}},\\
[15pt]
\displaystyle \frac{\partial\,\sigma_{xy}}{\partial\,t}-\mu\,\left( \frac{\partial\,v_{sy}}{\partial\,x}+\frac{\partial\,v_{sx}}{\partial\,y}\right) =f_{\sigma_{xy}},\\
[15pt]
\displaystyle \frac{\partial\,\sigma_{yy}}{\partial\,t}-\lambda_f\,\frac{\partial\,v_{sx}}{\partial\,x}-m\,\beta\,\frac{\partial\,w_x}{\partial\,x}- (\lambda_f + 2\mu)\,\frac{\partial\,v_{sy}}{\partial\,y} - m\,\beta \,\frac{\partial\,w_y}{\partial\,y}=f_{\sigma_{yy}},\\
[15pt]
\displaystyle \frac{\partial\,p}{\partial\,t} + m\, \left(  \beta\,\frac{\partial\,v_{sx}}{\partial\,x} +\frac{\partial\,w_x}{\partial\,x}+\beta\,\frac{\partial\,v_{sy}}{\partial\,y} +\frac{\partial\,w_y}{\partial\,y} \right)  =f_p,\\
[15pt]
\displaystyle \frac{\partial\,\psi _{xj}}{\partial\,t} + \frac{\rho _f}{\chi}\left( \,\frac{\partial\,\sigma_{xx}}{\partial\,x} + \frac{\partial\,\sigma_{xy}}{\partial\,y}\right)  + \frac{\rho }{\chi}\,\frac{\partial\,p}{\partial\,x} = \Omega \,w_x - \gamma\,\sum \limits _{\ell =1}^{N}a_{\ell}\, \psi _{x\ell} -(\theta_j + \Omega ) \,\psi _{xj}+f_{wx},\\
[15pt]
\displaystyle \frac{\partial\,\psi _{yj}}{\partial\,t} + \frac{\rho _f}{\chi}\,\left( \frac{\partial\,\sigma_{xy}}{\partial\,x} + \frac{\partial\,\sigma_{yy}}{\partial\,y}\right)  + \frac{\rho }{\chi}\,\frac{\partial\,p}{\partial\,y} = \Omega \,w_y - \gamma\,\sum \limits _{\ell =1}^{N}a_{\ell}\, \psi _{y\ell} -(\theta_j + \Omega ) \,\psi _{yj}+f_{wy}.
\end{array}
\right. 
\label{eq:syst2}
\end{equation}
where $j = 1,...,N$. Taking the vector of unknowns
\begin{equation}
{\bf U} = (v_{sx},v_{sy},w_x,w_y,\sigma_{xx},\sigma_{xy},\sigma_{yy},p,\psi_{x1},\psi_{y1},\ldots,\psi_{xN},\psi_{yN})^T
\label{eq:vect_U}
\end{equation}
and the source vector
\begin{equation}
{\bf F} = (f_{v_{xs}},f_{v_{ys}},\,f_{w_x},f_{w_y},\,f_{\sigma_{xx}},f_{\sigma_{xy}},f_{\sigma_{yy}},\,f_p,\,f_{w_x},f_{w_y},\ldots,\,f_{w_x},f_{w_y})^T,
\label{eq:vect_F}
\end{equation}
the system (\ref{eq:syst2}) can be written 
\begin{equation}
\frac{\partial\,{\bf U}}{\partial\,t} + {\bf A}\,\frac{\partial\,{\bf U}}{\partial\,x} + {\bf B}\,\frac{\partial\,{\bf U}}{\partial\,y} = -{\bf S}\,{\bf U} +{\bf F},
\label{eq:syst_hyperbolique}
\end{equation}
where ${\bf A}$ is the $(2\,N+8)^2$ propagation matrix along $x$, ${\bf B}$ is the $(2\,N+8)^2$ propagation matrix along $y$ and ${\bf S}$ is the $(2\,N+8)^2$ dissipation matrix (\ref{eq:matriceA},\,\ref{eq:matriceB},\,\ref{eq:matriceS}). The size of the system increases linearly with the number $N$ of diffusive variables.

%------------------------------------------------------------------------------------------

\subsection{Properties}\label{sec:DA:prop}

Four properties of system (\ref{eq:syst_hyperbolique}) are specified:
\begin{itemize}
\item the eigenvalues of ${\bf A}$ (\ref{eq:matriceA}) and ${\bf B}$ (\ref{eq:matriceB}) are real: $0$ with multiplicity $2\,N+2$, $\pm c_{pf}^{\infty}$, $\pm c_{ps}^{\infty}$ and $\pm c_{s}^{\infty}$. The system (\ref{eq:syst_hyperbolique}) is therefore hyperbolic;
\item since the eigenvalues and eigenvectors do not depend on the diffusive coefficients, they are the same in both the  Biot-DA and Biot-LF or Biot-JKD models. This is not so in the case of the method presented in the Ref. ~\onlinecite{MASSON10}, where the propagation matrix is modified to account for the fractional derivative;
\item the dispersion analysis is obtained in the case of the Biot-DA model by replacing $\widehat{F}$ by
\begin{equation}
\displaystyle \widehat{F}_{AD}(\omega) = \displaystyle \frac{\Omega + i\,\omega}{\sqrt{\Omega}}\,\sum \limits _{\ell=1}^{N}\frac{a_{\ell}}{\theta_{\ell} + \Omega + i\,\omega } 
\label{eq:fonction_correction_AD}
\end{equation}
in equations (\ref{eq:coef_relation_dis_P})-(\ref{eq:relation_dispersion_P}).
\end{itemize}

%------------------------------------------------------------------------------------------

\subsection{Determining the Biot-DA parameters}\label{sec:DA:coeff} 

The coefficients $a_{\ell}$ and $\theta_{\ell}$ in (\ref{eq:derivee_frac_rd}) have to be determined. For this purpose, we want to approach the original frequency correction $\widehat{F}_{JKD}(\omega)$ (\ref{eq:fonction_correction}) by $\widehat{F}_{AD}(\omega)$ (\ref{eq:fonction_correction_AD}) in a given frequency range of interest. Let $Q(\omega)$ be the optimized quantity and $Q_{ref}(\omega)$ be the desired quantity
\begin{equation}
\left\lbrace 
\begin{array}{ll}
\displaystyle Q(\omega) & = \displaystyle \frac{\widehat{F}_{AD}(\omega)}{\widehat{F}_{JKD}(\omega)} = \sum \limits _{\ell=1}^{N}a_{\ell}\,\frac{(\Omega + i\,\omega)^{1/2}}{\theta_{\ell} + \Omega + i\,\omega } = \sum \limits _{\ell=1}^{N}a_{\ell}\,q_{\ell}(\omega),\\
[15pt]
\displaystyle Q_{ref}(\omega) & = 1.
\end{array}
\right. 
\label{eq:q_opti}
\end{equation}
We implement a linear optimization procedure\cite{EMMERICH87,GROBY06,LOMBARD11,BLANC12} in order to minimize the distance between $Q(\omega)$ and $Q_{ref}(\omega)$ in the interval $[\omega_{min},\omega_{max}]$ centered on $\omega_0 = 2\,\pi\,f_0$, where $f_0$ is the central frequency of the source. The abscissas $\theta_{\ell}$ are fixed and distributed linearly on a logarithmic scale
\begin{equation}
\theta_{\ell} = \omega_{min}\left( \frac{\omega_{max}}{\omega_{min}}\right) ^{\frac{\ell-1}{N-1}}\mbox{,}\qquad \ell = 1,...,N.
\end{equation}
The weights $a_{\ell}$ are obtained by solving the system
\begin{equation}
\sum \limits _{\ell=1}^{N}a_{\ell}\,q_{\ell}(\tilde{\omega}_k) = 1\mbox{,}\qquad k = 1,...,K,
\end{equation}
where the $\tilde{\omega}_k$ are also distributed linearly on a logarithmic scale of $K$ points
\begin{equation}
\tilde{\omega}_k = \omega_{min}\left( \frac{\omega_{max}}{\omega_{min}}\right) ^{\frac{k-1}{K-1}}\mbox{,}\qquad k = 1,...,K.
\label{eq:omega_k}
\end{equation}
Since the $q_{\ell}(\omega)$ are complex functions, optimization is performed simultaneously on the real and imaginary parts 
\begin{equation}
\left\lbrace 
\begin{array}{ll}
\displaystyle \sum \limits _{\ell=1}^{N}a_{\ell}\,\mathbb{R}\mbox{e}(q_{\ell}(\tilde{\omega}_k))  & = 1,\\
[12pt]
\displaystyle \sum \limits _{\ell=1}^{N}a_{\ell}\,\mathbb{I}\mbox{m}(q_{\ell}(\tilde{\omega}_k))  & = 0\mbox{,}\qquad k=1,...,K.\\
\end{array}
\right. 
\label{eq:syst_opti}
\end{equation}
For practical purposes, we use $\omega_{min}=\omega_0/10$ and $\omega_{max}=10\,\omega_0$, and we solve an overdetermined system ($K=N$), which can be solved by writing normal equations.\cite{NRPAS} The number of diffusive variables $N$ can be determined in terms of the ratio $f_0/f_c$ and of the chosen accuracy: see Ref. ~\onlinecite{BLANC12} for details.

%------------------------------------------------------------------------------------------
%------------------------------------------------------------------------------------------

\section{Numerical modeling}\label{sec:num}

\subsection{Splitting}\label{sec:num:splitting}

In order to integrate the Biot-DA system (\ref{eq:syst_hyperbolique}), a uniform grid is introduced, with mesh size $\Delta\,x = \Delta\,y$ and time step $\Delta\,t$. The approximation of the exact solution ${\bf U}(x_i = i\,\Delta\,x,y_j = j\,\Delta\,y,t_n = n\,\Delta\,t)$ is denoted by ${\bf U}_{ij}^n$. If an unsplit integration of (\ref{eq:syst_hyperbolique}) is performed, a Von-Neumann analysis typically yields the stability condition
\begin{equation}
\Delta\,t \leq \min \left( \Upsilon\,\frac{\Delta\,x}{c_{pf}^{\infty}} \; , \; \frac{2}{R({\bf S})}\right) ,
\label{eq:CFL_direct}
\end{equation}
where $R({\bf S})$ is the spectral radius of ${\bf S}$, and $\Upsilon>0$ depends on the numerical scheme. We have no theoretical estimate of $R({\bf S})$, but numerical studies have shown that this value is similar to that of the spectral radius in LF: $\frac{\eta}{\kappa}\,\frac{\rho}{\chi}$, which can be very large.\cite{CHIAVASSA11} The time step can therefore be highly penalized in this case (\ref{eq:CFL_direct}).

A more efficient strategy is adopted here, which consists in splitting the original system (\ref{eq:syst_hyperbolique}) into a propagative part and a diffusive part:
\begin{equation}
\left\lbrace 
\begin{array}{l}
\displaystyle
\frac{\partial\,{\bf U}}{\partial\,t}+{\bf A}\,\frac{\partial\,{\bf U}}{\partial\,x}+{\bf B}\,\frac{\partial\,{\bf U}}{\partial\,y} = 0,\\
[12pt]
\displaystyle
\frac{\partial\,{\bf U}}{\partial\,t} = -{\bf S}\,{\bf U}.
\end{array}
\right. 
\label{eq:splitting}
\end{equation}
For the sake of simplicity, the source term ${\bf F}$ has been omitted here. The discrete operators associated with the propagative part and the diffusive part are denoted by ${\bf H}_a$ and ${\bf H}_b$, respectively. The second-order Strang splitting is then used to integrate (\ref{eq:syst_hyperbolique}) between $t_n$ ant $t_{n+1}$, giving the  time-marching
\begin{equation}
\begin{array}{lllll}
\displaystyle
&\bullet& {\bf U}_{ij}^{(1)}&=&{\bf H}_{b}(\frac{\Delta\,t}{2})\,{\bf U}_{ij}^{n},\\
[6pt]
\displaystyle
&\bullet& {\bf U}_{ij}^{(2)}&=&{\bf H}_{a}(\Delta\,t)\,{\bf U}_{ij}^{(1)},\\
[6pt]
\displaystyle
&\bullet& {\bf U}_{ij}^{n+1}&=&{\bf H}_{b}(\frac{\Delta\,t}{2})\,{\bf U}_{ij}^{(2)}.
\end{array}
\label{eq:algo_splitting}
\end{equation}
The discrete operator ${\bf H}_a$ associated with the propagative part (first equation of (\ref{eq:splitting})) is an ADER 4 (Arbitrary DERivatives) scheme.\cite{SCHWARTZKOPFF04} This scheme is fourth-order accurate in space and time, is dispersive of order 4 and dissipative of order 6, and has a stability limit $\Upsilon=1$. On Cartesian grids, ADER 4 amounts to a fourth-order Lax-Wendroff scheme.\\

Since the physical parameters do not vary with time, the diffusive part  (second equation of \ref{eq:splitting})) can be solved exactly. This gives
\begin{equation}
{\bf H}_b\left(\frac{\Delta\,t}{2}\right)\,{\bf U}_{ij} = e^{-\frac{\Delta\,t}{2}\,{\bf S}}\,{\bf U}_{ij}.
\label{eq:split_diffu_exp}
\end{equation}
The matrix $e^{-\frac{\Delta\,t}{2}\,{\bf S}}$ is computed numerically using the $(6,6)$ Pad\'e approximation in the "scaling and squaring method".\cite{MOLER03} The numerical integration of the diffusive step (\ref{eq:split_diffu_exp}) is unconditionally stable.\cite{BLANC12}\\

The full algorithm (\ref{eq:algo_splitting}) is therefore stable under the optimum CFL (Courant-Friedrichs-Lewy) stability condition
\begin{equation}
\mbox{CFL} = c_{pf}^{\infty}\,\frac{\Delta\,t}{\Delta\,x} \leqslant \Upsilon,
\label{eq:CFL_dsplit}
\end{equation}
which is always independent of the Biot-DA model coefficients.

%------------------------------------------------------------------------------------------

\subsection{Immersed interface method}\label{sec:num:IIM}

\begin{figure}[htbp]
\begin{center}
\begin{tabular}{c}
\includegraphics[scale=0.7]{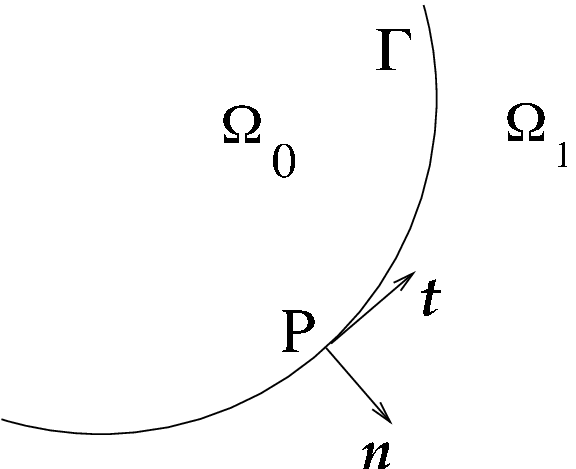}
\end{tabular}
\end{center}
\caption{Interface $\Gamma$ between two poroelastic media $\Omega_0$ and $\Omega_1$.}
\label{fig:interface}
\end{figure}

Let us consider two homogeneous poroelastic media $\Omega_0$ and $\Omega_1$ separated by a stationary interface $\Gamma$, as shown in figure \ref{fig:interface}. The governing equations (\ref{eq:syst2}) in each medium have to be completed by a set of jump conditions. The simple case of perfect bonding and perfect hydraulic contact along $\Gamma$ is considered here, modeled by the jump conditions:\cite {GUREVICH99}
\begin{equation}
[{\bf v}_s] = {\bf 0},\quad [{\bf w}.{\bf n}] = 0,\quad [\sigma.{\bf n}] = {\bf 0},\quad [p] = 0.
\label{eq:jump_condition}
\end{equation}
The discretization of the jump conditions requires special care, for three reasons. First, if the interfaces do not coincide with the uniform meshing, then geometrical errors will occur. Secondly, the jump conditions (\ref{eq:jump_condition}) will not be enforced numerically by the finite-difference scheme, and the numerical solution will therefore not converge towards the exact solution. Lastly, the smoothness of the solution required to solve (\ref{eq:syst_hyperbolique}) will not be satisfied across the interface, which will decrease the convergence rate of the ADER scheme.

To overcome these drawbacks without detracting the efficiency of Cartesian grid methods, an immersed interface method\cite{CHIAVASSA11,CHIAVASSA12} is used. The latter studies can be consulted for a detailed description of this method. The basic principle is as follows:  at the irregular nodes where the ADER scheme crosses an interface, modified values of the solution are used on the other side of the interface instead of the usual numerical values.

Calculating these modified values is a complex task involving high-order derivation of jump conditions (\ref{eq:jump_condition}), Beltrami-Michell relations (\ref{eq:beltrami}) and singular value decompositions. Fortunately, all these time consuming procedures can be carried out during a preprocessing stage and only small matrix-vector multiplications need to be performed during the simulation. After optimizing the code, the extra CPU cost can be practically negligible, i.e. lower than 1\% of that required by the time-marching procedure.

The immersed interface method uses only the propagative part (first equation of (\ref{eq:splitting})) and the jump conditions (\ref{eq:jump_condition}), whose do not depend on the frequency regime. Consequently, no changes are required in the high-frequency range compared to the low-frequency range. The algorithm is the same as in the low-frequency range, see Ref. ~\onlinecite{CHIAVASSA11} for details.

%------------------------------------------------------------------------------------------
%------------------------------------------------------------------------------------------

\section{Numerical experiments}\label{sec:exp}

\noindent
{\it General configuration}\label{sec:exp:config}

\begin{table}[h!]
\begin{center}
\begin{tabular}{llll}
 & Parameters $\qquad$ & $\Omega_0$ $\qquad\qquad$ & $\Omega_1$\\
\hline
\rule[-1mm]{0mm}{5mm} Saturating fluid & $\rho_f$ (kg/m$^3$) & $1040$ & $1000$\\
\rule[-1mm]{0mm}{5mm} & $\eta$ (Pa.s)  & $1.5\,10^{-3}$ & $10^{-3}$\\
\rule[-1mm]{0mm}{5mm} Grain & $\rho_s$ (kg/m$^3$) & $2650$ & $2644$\\
\rule[-1mm]{0mm}{5mm} & $\mu$ (Pa) & $2.93\,10^{9}$ & $7.04\,10^{9}$\\
\rule[-1mm]{0mm}{5mm} Matrix & $\phi$ & $0.335$ & $0.2$\\
\rule[-1mm]{0mm}{5mm} & $a$ & $2$ & $2.4$\\
\rule[-1mm]{0mm}{5mm} & $\kappa$ (m$^2$) & $10^{-11}$ & $3.6\,10^{-13}$\\
\rule[-1mm]{0mm}{5mm} & $\lambda_f$ (Pa) & $6.14\,10^{9}$ & $1.06\,10^{10}$\\
\rule[-1mm]{0mm}{5mm} & $m$ (Pa) & $6.49\,10^{9}$ & $9.70\,10^{9}$\\
\rule[-1mm]{0mm}{5mm} & $\beta$ & $9.56\,10^{-1}$ & $7.20\,10^{-1}$\\
\rule[-1mm]{0mm}{5mm} & $\Lambda$ (m) & $2.19\,10^{-5}$ & $5.88\,10^{-6}$\\ \hline
\rule[-1mm]{0mm}{5mm} Phase velocities & $c_{pf}^{\infty}$ (m/s) & $2384.17$ & $3269.89$\\
\rule[-1mm]{0mm}{6mm} & $c_{ps}^{\infty}$ (m/s) & $758.95$ & $814.95$\\
\rule[-1mm]{0mm}{6mm} & $c_{s}^{\infty}$ (m/s) & $1229.00$ & $1776.16$\\
\rule[-1mm]{0mm}{6mm} & $f_c$ (Hz) & $3.84\,10^{3}$ & $3.68\,10^{4}$\\ \hline
\end{tabular}
\end{center}
\caption{Physical parameters used in numerical experiments.}
\label{table:para_phy}
\end{table}

The physical parameters used in all the numerical experiments, which are given in table \ref{table:para_phy}, correspond to Cold Lake sandstone ($\Omega_0$) and Berea sandstone ($\Omega_1$) saturated with water.
The computational domain $[-0.1,0.1]^2$ m is discretized with $N_x\times N_y$ grid points, $N_x = N_y = 1600$, which amounts to 29 points per slow compressional wavelength in $\Omega_0$ and in $\Omega_1$ for a source of central frequency $f_0 = 200$ kHz. As seen in 1D,\cite{BLANC12} $N=6$ diffusive variables are used to ensure an error of model smaller than $5.58$ \%. The time step is deduced from (\ref{eq:CFL_dsplit}), taking  CFL $= 0.95$. The numerical experiments are performed on an Intel Core i7 processor at $2.80$ GHz.

%------------------------------------------------------------------------------------------
\newpage
\noindent
{\it Test 1: homogeneous medium}\label{sec:exp:test1}

\begin{figure}[h!]
\begin{center}
\begin{tabular}{cc}
$p$ & $\sigma_{yy}$\\
\includegraphics[scale=0.45]{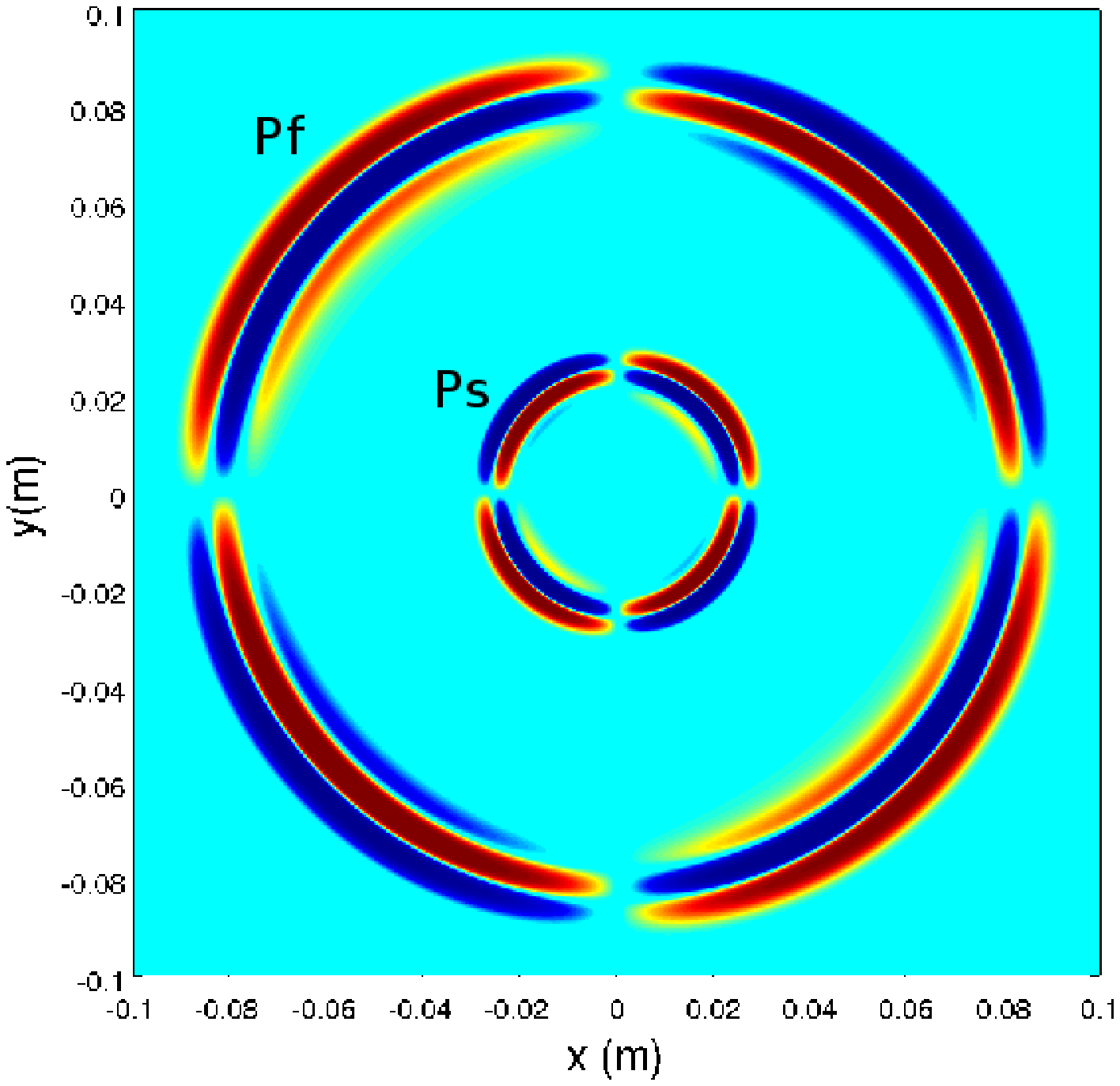} & 
\includegraphics[scale=0.45]{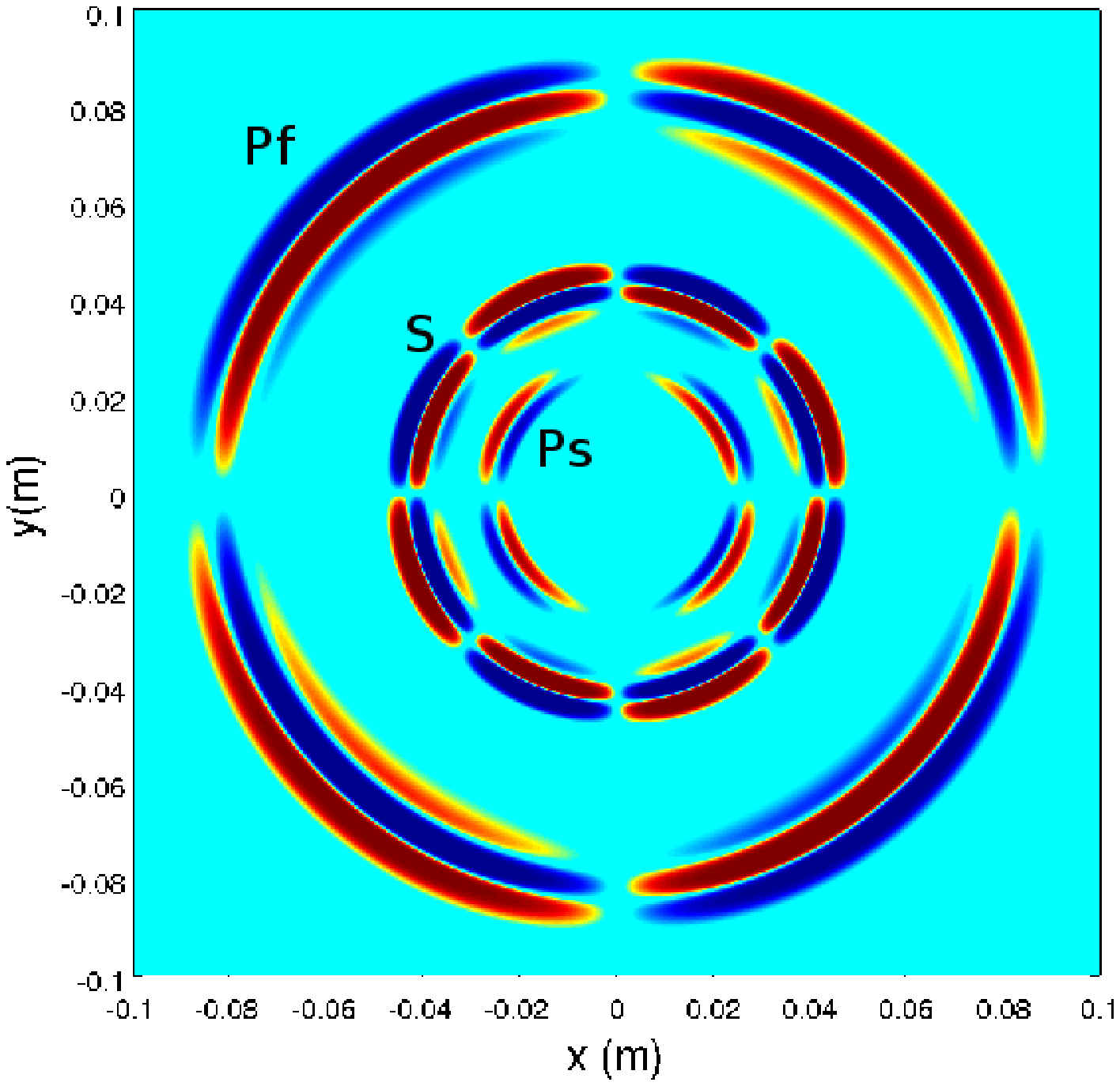}
\end{tabular}
\end{center}
\caption{Test 1. Fast and slow compressional waves, respectively $P_f$ and $P_s$, and shear wave $S$ emitted by a source point at $(0.01,0)$. Pressure (left) and $\sigma_{yy}$ component of the stress tensor (right) at $t_1 = 3.98\,10^{-5}$ s.}
\label{fig:test1}
\end{figure}

The unbounded homogeneous medium $\Omega_0$ is excited by a point source. The only non-null component of the vector of source $\bm{F}$ (\ref{eq:vect_F}) is $f_{\sigma_{xy}}=g(x,y)\,h(t)$, where $g(x,y)$ is a truncated gaussian centered at point $(0,0)$, of radius $R_0=3.79\,10^{-3}$ m and $\Sigma=1.90\,10^{-3}$ m:
\begin{equation}
g(x,y) =
\left\lbrace 
\begin{array}{ll}
\displaystyle \frac{1}{\pi\,\Sigma^2}\,\exp\left(-\frac{x^2+y^2}{\Sigma^2}\right) \quad & \mbox{if}\;0\leqslant x^2+y^2\leqslant R_0^2,\\
[10pt]
\displaystyle 0 & \mbox{otherwise},
\end{array}
\right. 
\label{eq:gaussienne}
\end{equation}
and $h(t)$ is a Ricker signal of central frequency $f_0 = 200$ kHz and of time-shift $t_0=2/f_0=10^{-5}$ s:
\begin{equation}
h(t) =
\left\lbrace 
\begin{array}{ll}
\displaystyle \left( 2\,\pi^2\,f_0^2\,\left( t-\frac{1}{f_0}\right) ^2-1\right) \,\exp\left( -\pi^2\,f_0^2\,(t-\frac{1}{f_0})^2\right) \quad & \displaystyle \mbox{if}\;0\leqslant t\leqslant t_0,\\
[10pt]
\displaystyle 0 & \displaystyle \mbox{otherwise}.
\end{array}
\right. 
\label{eq:ricker}
\end{equation}
We use a truncated gaussian for $g(x,y)$ rather than a Dirac to avoid spurious numerical artifacts localized around the source point. This source generates cylindrical waves of all types: fast and slow compressional waves and shear wave, which are denoted $P_f$, $P_s$ and $S$, respectively, in figure \ref{fig:test1}. Fast and slow compressional waves are observed as regards the pressure, while the additional shear wave is present in the $\sigma_{yy}$ component of the stress tensor. No special care is applied to simulate outgoing waves (with PML, for instance), since the simulation is stopped before the waves have reached the edges of the computational domain.\\

%------------------------------------------------------------------------------------------

\noindent
{\it Test 2: diffraction of a plane wave by a plane interface}\label{sec:exp:test2}

In all the following tests, the source is a plane right-going fast compressional wave, whose wavevector $\bm{k}$ makes an angle $\theta=0$ with the horizontal $x$-axis. Its time evolution is the same Ricker signal as in the first test (\ref{eq:ricker}). We use periodic boundary conditions at the top and at the bottom of the domain.\\

\begin{figure}[h!]
\begin{center}
\begin{tabular}{cc}
(a) & (b)\\
\includegraphics[scale=0.45]{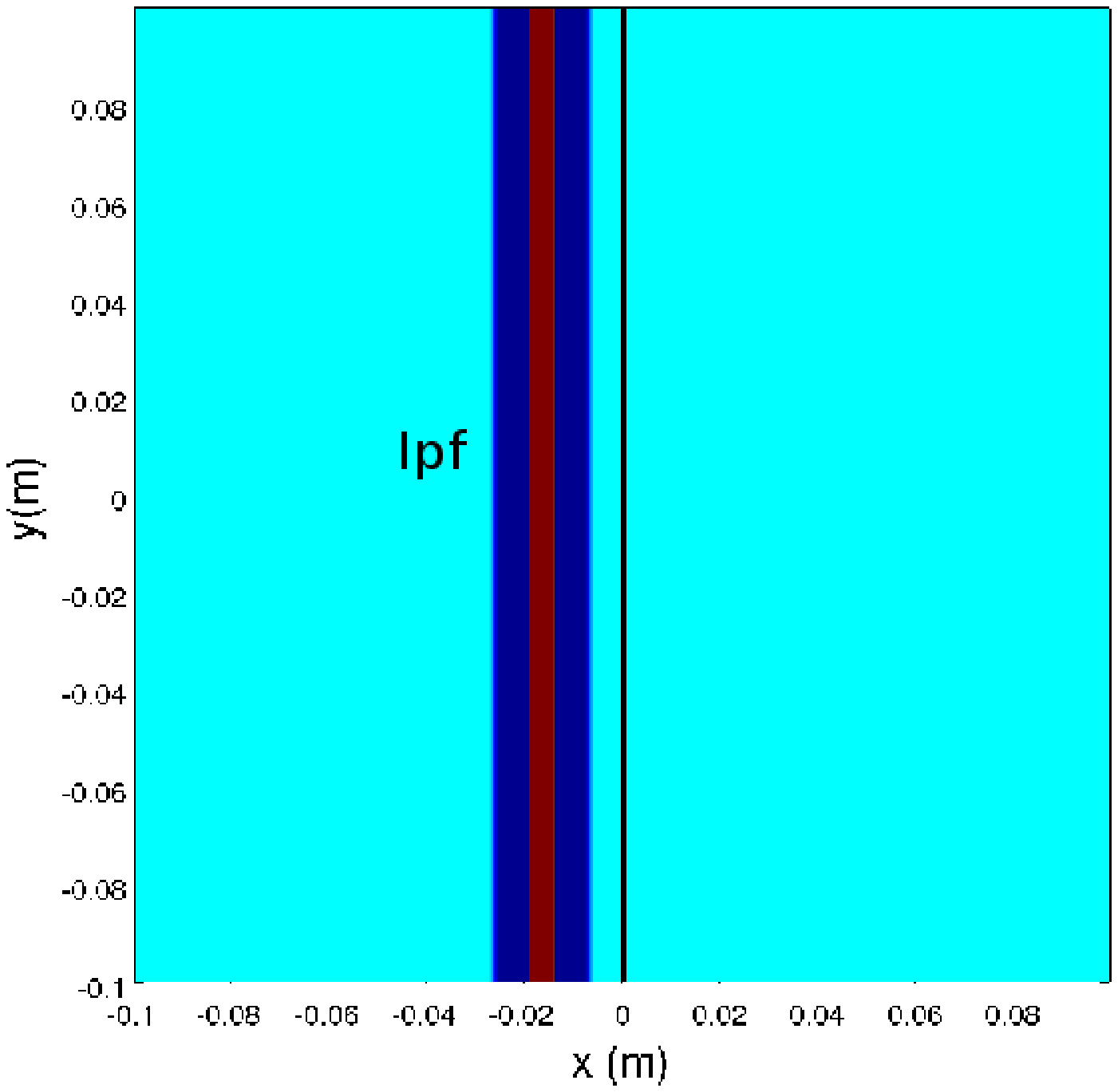} & 
\includegraphics[scale=0.45]{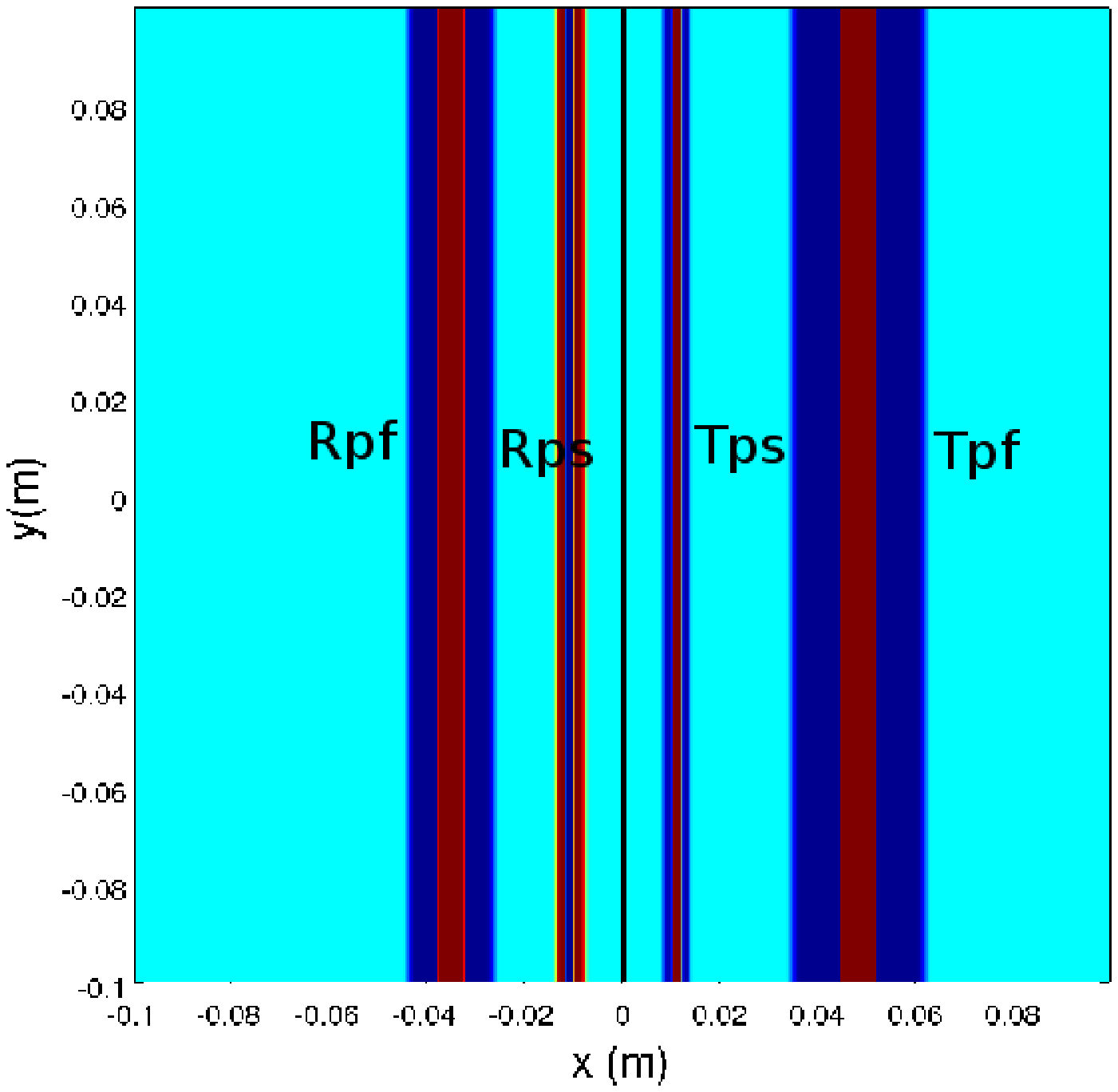}\\
\end{tabular}
\end{center}
\caption{Test 2. Snapshot of $p$ at initial time (a) and at $t_1=1.82\,10^{-5}$ s (b).}
\label{fig:test2_carte}
\end{figure}

This test is conducted on a vertical plane interface at $x=0$ m. The incident $P_f\,$-wave (Ip$_f$) propagates in the medium $\Omega_0$. The incident wave crosses the interface normally, leading to a 1D configuration. Consequently, only the fast and slow compressional waves propagate. From a numerical point of view, however, the problem is fully bidimensional. The advantage of such a 1D configuration is that each diffracted wave has interacted with the interface and is consequently very sensitive to the discretization of the interface conditions (\ref{eq:jump_condition}). The figure \ref{fig:test2_carte} shows a snapshot of the pressure at $t_1=1.82\,10^{-5}$ s, on the whole computational domain. The reflected fast and slow compressional waves, denoted respectively Rp$_f$ and Rp$_s$, propagate in the medium $\Omega_0$; and the transmitted fast and slow compressional  waves, denoted respectively Tp$_f$ and Tp$_s$, propagate in the medium $\Omega_1$.\\

\begin{figure}[h!]
\begin{center}
\begin{tabular}{cc}
Pressure & Zoom on the slow compressional waves\\
\includegraphics[scale=0.45]{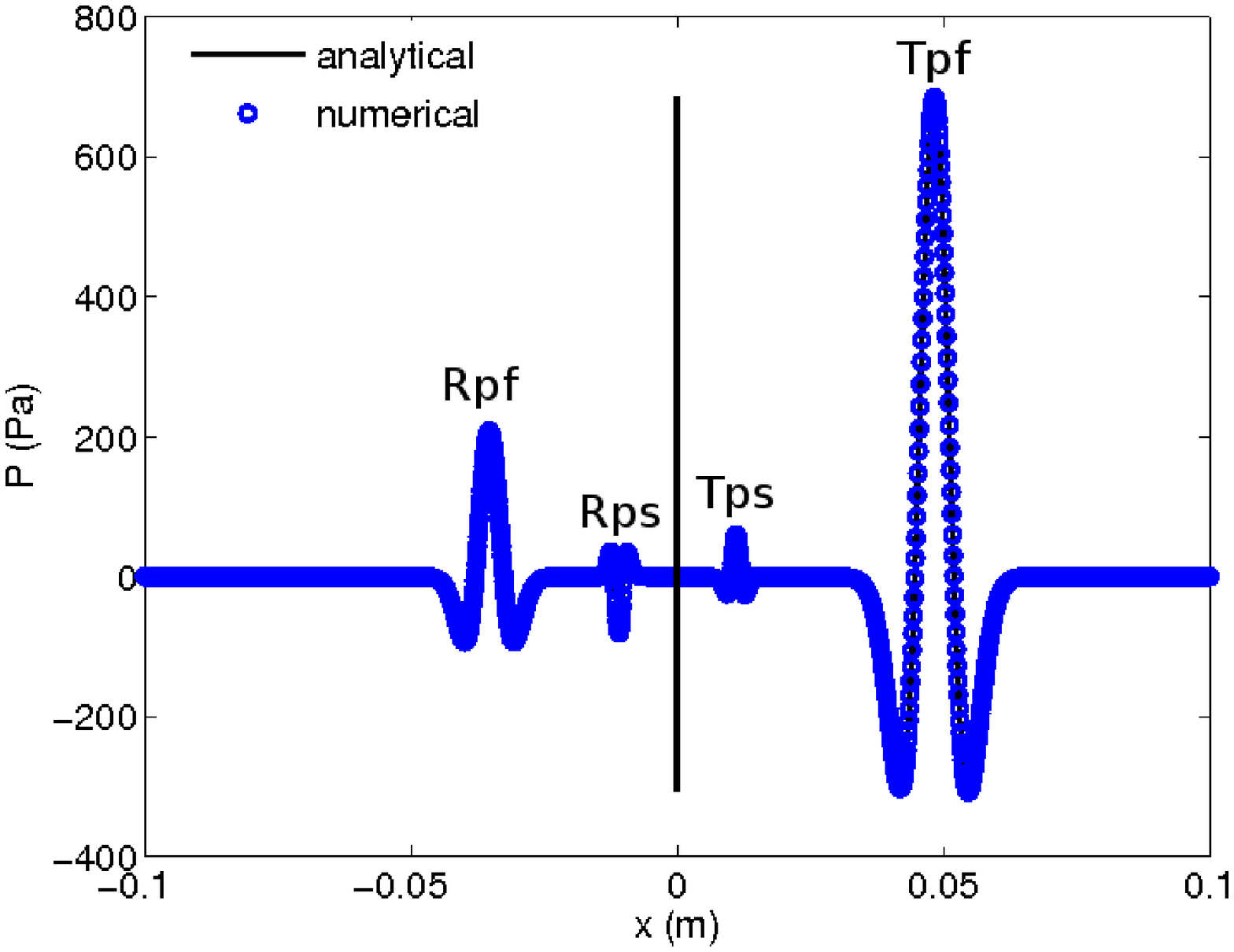} & 
\includegraphics[scale=0.45]{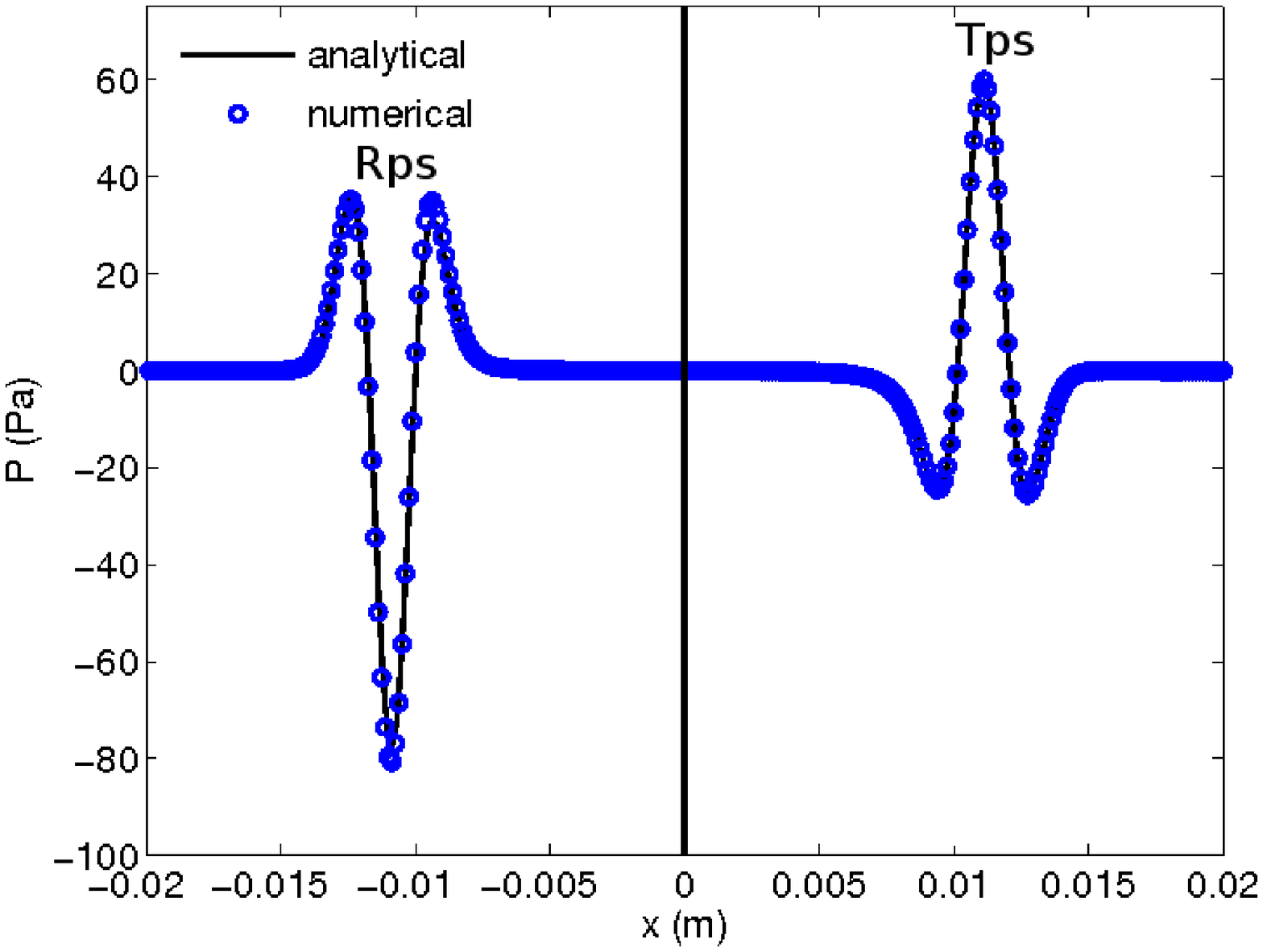}\\
\end{tabular}
\end{center}
\caption{Test 2. Pressure along the line $y=0$ m; vertical line denote the interface. Comparison between the numerical values (circle) and the analytical values (solid line) of $p$ at $t_1=1.82\,10^{-5}$ s.}
\label{fig:test2_coupe}
\end{figure}

In this case, we compute the exact solution of Biot-DA thanks to standard tools of Fourier analysis. Special care is taken to ensure that the initial data and the exact solution are valuable reference solutions : $N_f=512$ Fourier modes are used, with a frequency step $\Delta\,f=4000$ Hz. The figure \ref{fig:test2_coupe} shows the excellent agreement between the analytical and the numerical values of the pressure along the line $y=0$ m.\\

%------------------------------------------------------------------------------------------

\noindent
{\it Test 3: diffraction of a plane wave by a cylinder or a circular shell}\label{sec:exp:test3}

\begin{figure}[h!]
\begin{center}
\begin{tabular}{cc}
(a) & (b)\\
\includegraphics[scale=0.45]{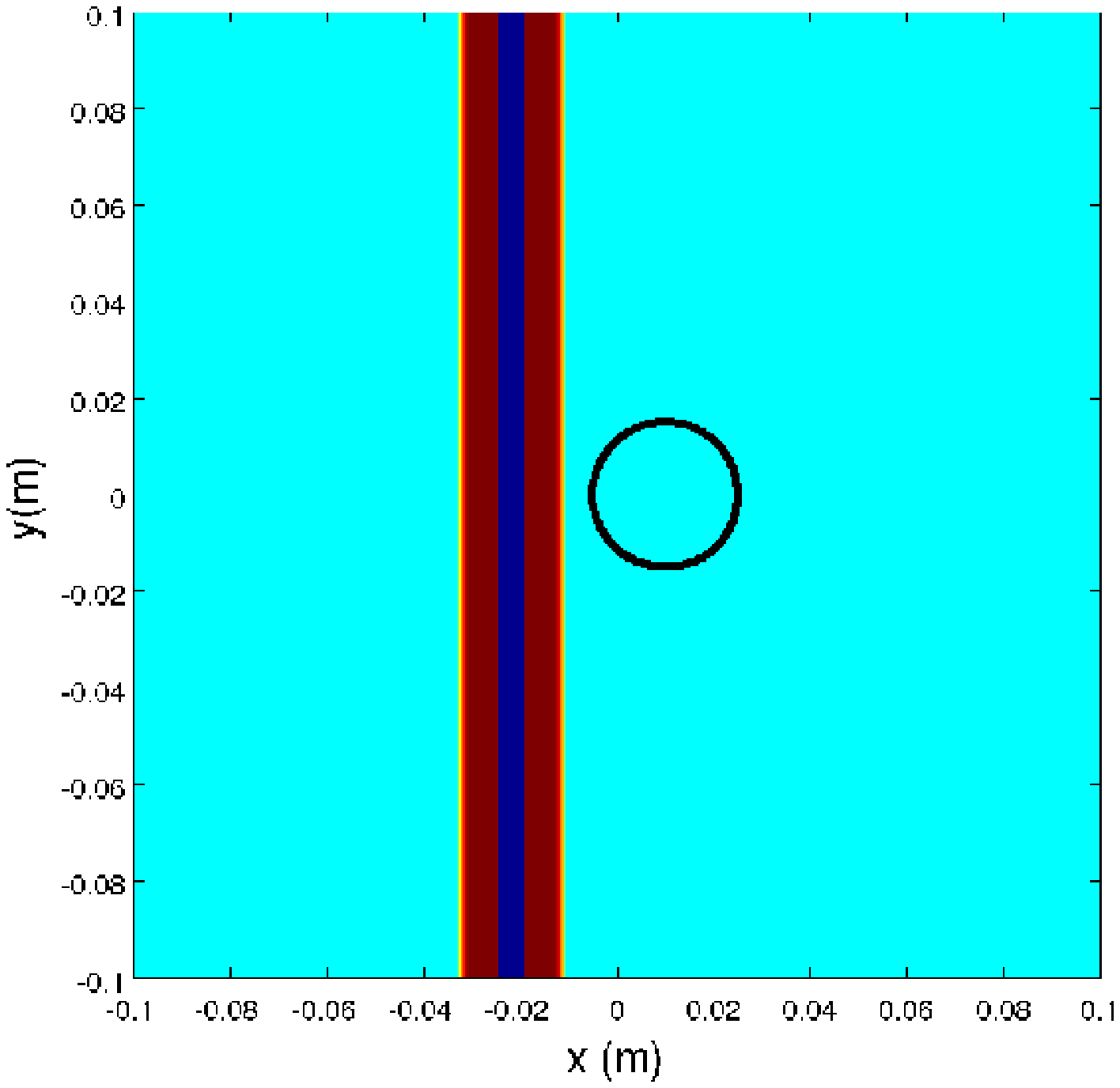} & 
\includegraphics[scale=0.45]{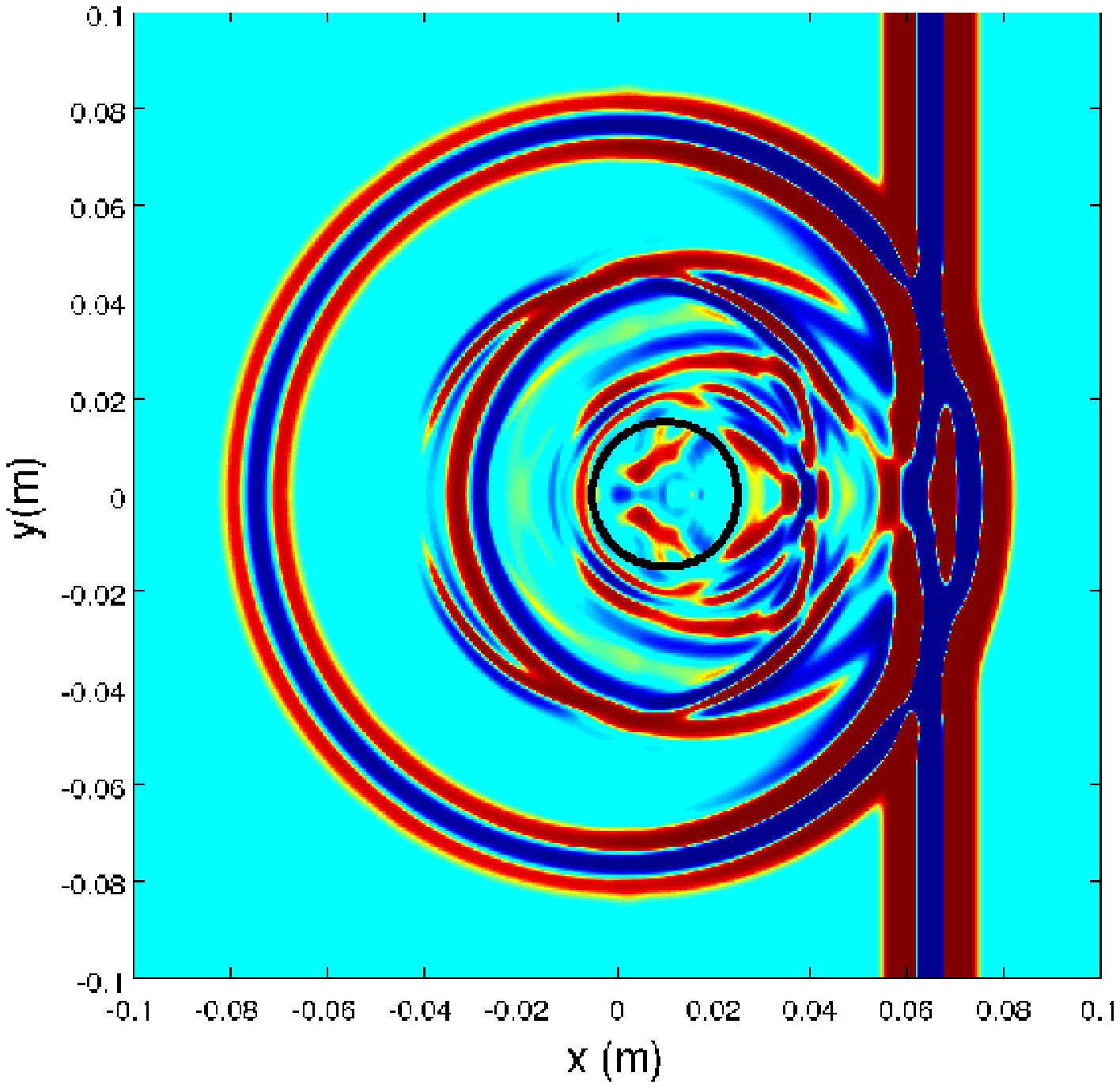}\\
\end{tabular}
\end{center}
\caption{Test 3. Scatterer: cylinder. Snapshot of $\sigma_{yy}$ at initial time (a) and at $t_1 = 3.63\,10^{-5}$ s (b).}
\label{fig:test3_cercle}
\end{figure}

\begin{figure}[h!]
\begin{center}
\begin{tabular}{cc}
(a) & (b)\\
\includegraphics[scale=0.45]{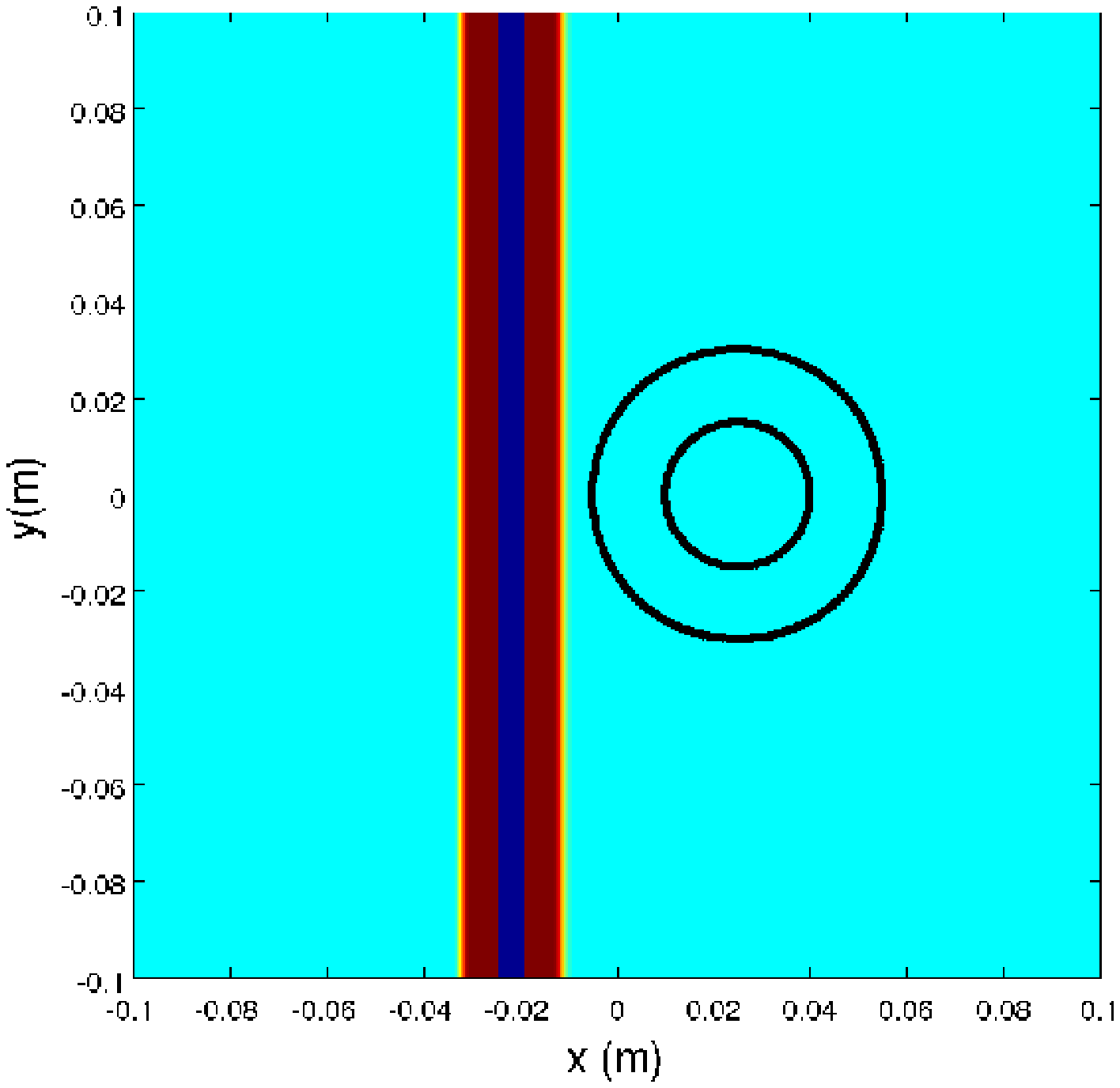} & 
\includegraphics[scale=0.45]{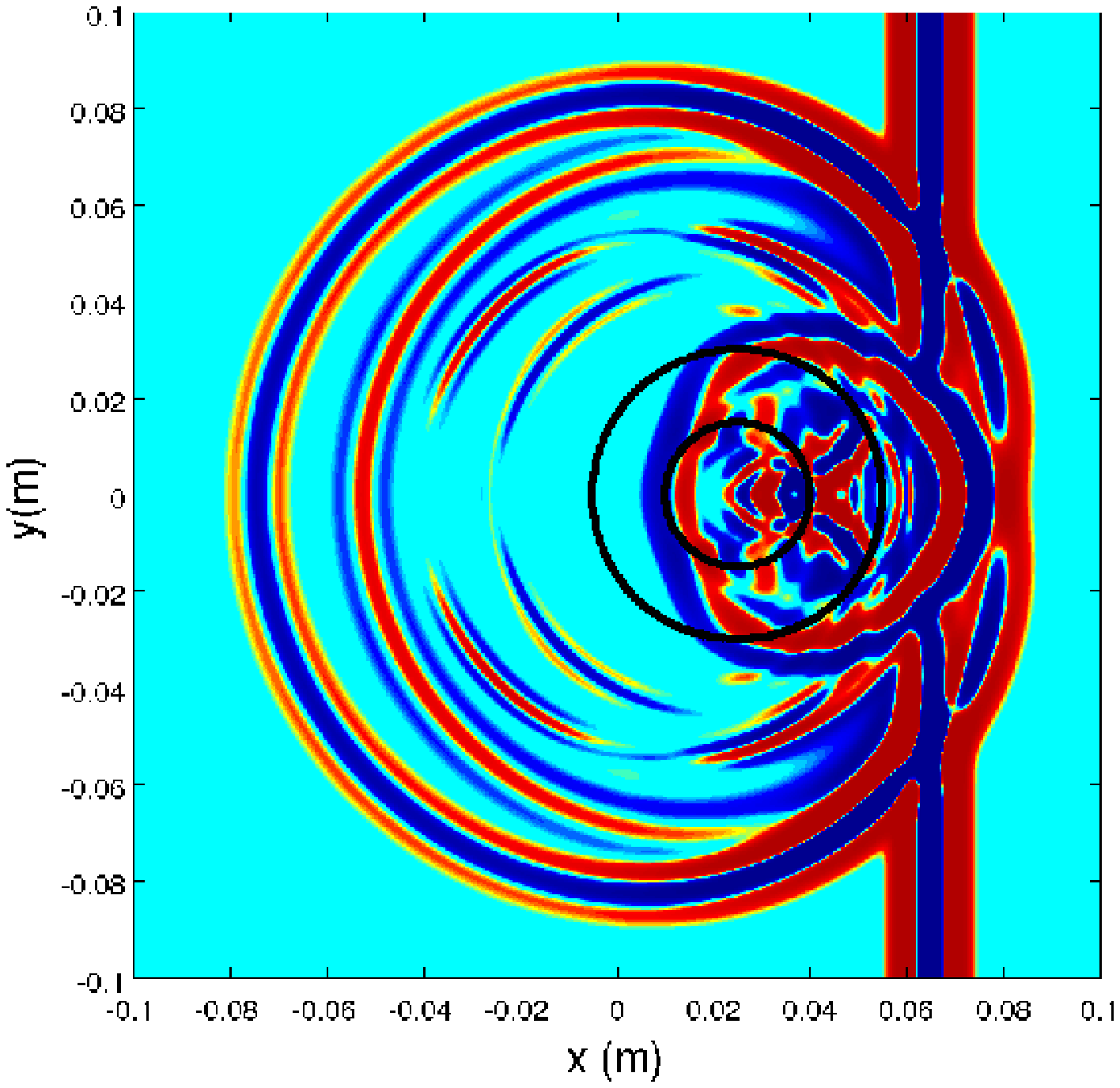}\\
\end{tabular}
\end{center}
\caption{Test 3. Scatterer: circular shell. Snapshot of $\sigma_{yy}$ at initial time (a) and at $t_1 = 3.63\,10^{-5}$ s (b).}
\label{fig:test3_couronne}
\end{figure}

In the first case, a cylindrical scatterer $\Omega_1$ of radius $0.015$ m is centered at point $(0.01,0)$. In the second case, we consider a poroelastic shell $\Omega_1$ of external radius $0.03$ m and internal radius $0.015$ m, centered at point $(0.025,0)$. The initial conditions are illustrated in figure \ref{fig:test3_cercle}(a) and \ref{fig:test3_couronne}(a), and the figure \ref{fig:test3_cercle}(b) and \ref{fig:test3_couronne}(b) represent a snapshot of $\sigma_{yy}$ at $t_1 = 3.63\,10^{-5}$ s.
The immersed interface method ensures that no spurious diffractions are created during the interaction of the incident wave with the scatterers, despite the non-null curvature of the interfaces. In the figure \ref{fig:test3_cercle}(b) and \ref{fig:test3_couronne}(b), the transmitted fast compressional wave has a curved wavefront because its phase velocity $c_{pf}$ in the medium $\Omega_1$ is greater than in the medium $\Omega_0$. Classical conversions and scattering phenomena are observed.

%------------------------------------------------------------------------------------------
\noindent
{\it Test 4: multiple ellipsoidal scatterers}\label{sec:exp:test5}

In the last test, the ability of the proposed numerical strategy to handle complex geometries with variable curvature is illustrated. $15$ ellipsoidal scatterers, of major radius $0.008$ m and of minor radius $0.005$ m, of medium $\Omega_1$ are randomly embedded in $\Omega_0$. Performing such a simulation with hundreds of scatterers has physical applications. Doing so provide a mean to check the validity of multiple scattering models.\cite{LUPPE08,CHEKROUN12}\\
%The effective field, which is obtained by averaging the fields in all the possible disordered configurations, corresponds to that of waves propagating in an effective homogeneous medium.

\begin{figure}[h!]
\begin{center}
\begin{tabular}{cc}
(a) & (b)\\
\includegraphics[scale=0.45]{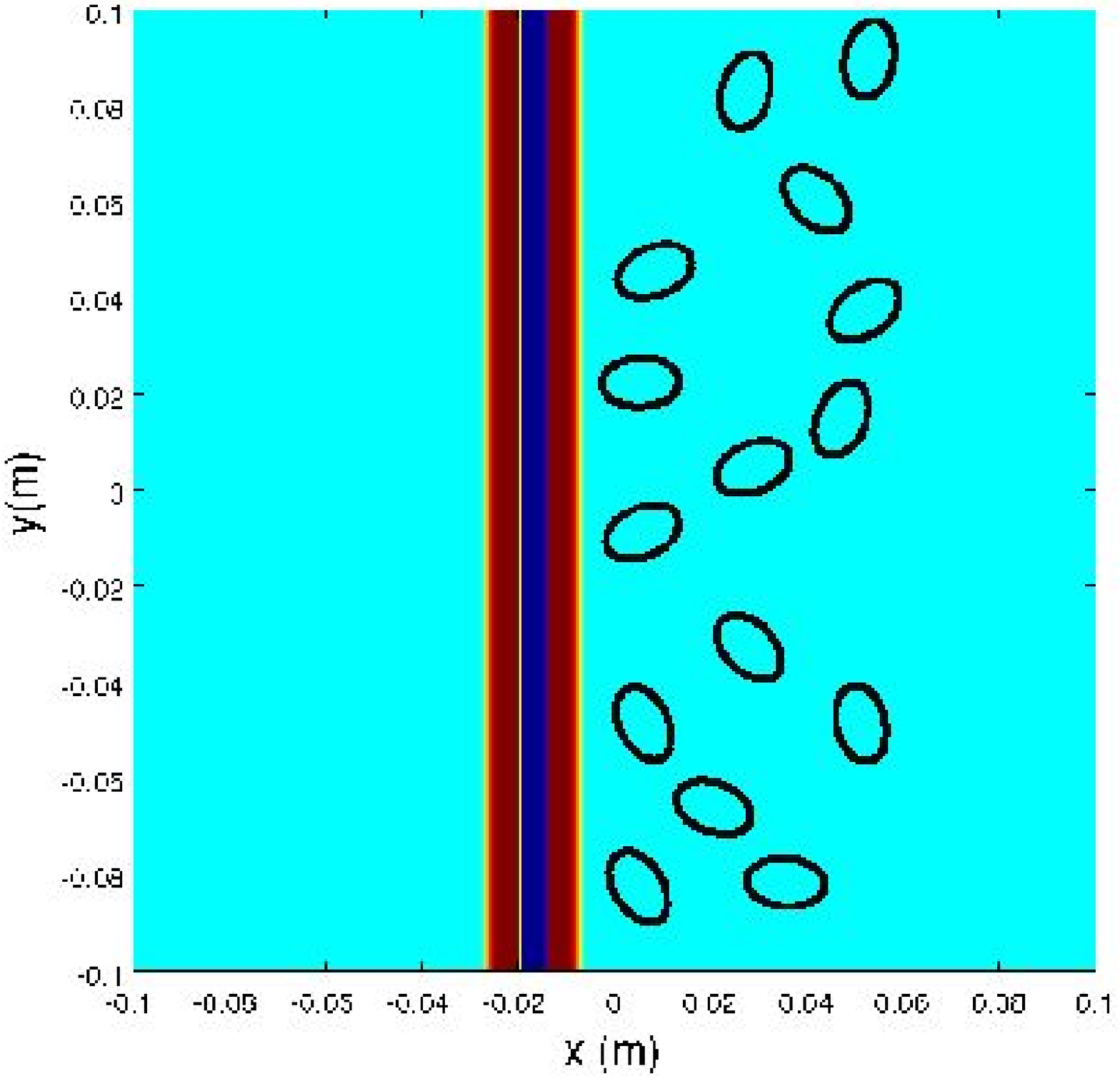} & 
\includegraphics[scale=0.45]{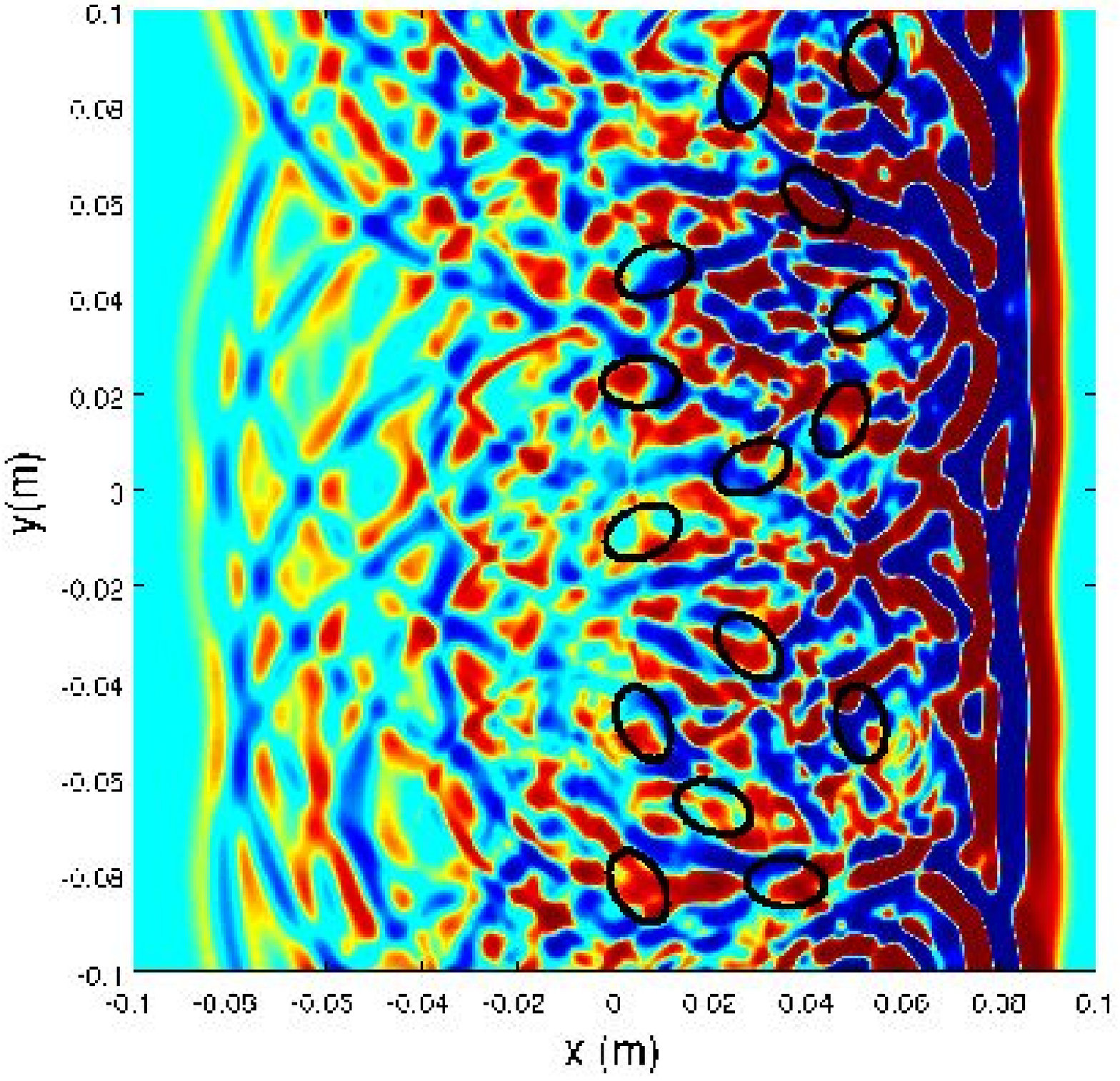}\\
\end{tabular}
\end{center}
\caption{Test 4. Snapshot of $\sigma_{yy}$ at initial time (a) and at $t_1=4.03\,10^{-5}$ s (b).}
\label{fig:test4}
\end{figure}

%------------------------------------------------------------------------------------------
%------------------------------------------------------------------------------------------
\section{Conclusion}\label{sec:conclu}

An explicit finite-difference method is presented here for simulating transient poroelastic waves in the full range of validity of poroelasticity. The Biot-JKD model, which involves order 1/2 fractional derivatives, was replaced here by an approximate Biot-DA model, which is much more tractable numerically. The Biot-DA coefficients are determined here using an optimization procedure, which depends on the frequency range of interest. The hyperbolic system of partial differential equations was discretized using efficient tools (Strang splitting, fourth-order ADER scheme, immersed interface method). This numerical methodology  provides highly accurate numerical solutions, allowing a fine numerical investigation of realistic porous media. 

Some future lines of research are suggested:
\begin{itemize}
\item \emph{Continuously variable porous media.} In the previous numerical experiments, the physical parameter characterizing the porous media are constant and discontinuous across the interfaces. The possibility of applying the presented numerical method to 2D  porous media with continuously variable coefficients,\cite{GROBY11} where no analytical expressions are available, is currently in progress: see Ref. ~\onlinecite{BLANC12} in 1D. 
\item \emph{Multiple scattering.} The numerical tools presented here make possible the modeling of multiple scattering in random media. Based on simulated data, the properties of the {\it effective medium} amount to the disordered under study can be deduced.\cite{LUPPE08,CHEKROUN12} Optimization and parallelization of the method must be realized to take into account of hundreds of inclusions in our simulations.
\item \emph{Anisotropy.} The anisotropy of some porous media is sometimes a main feature, for instance in biomechanics to model trabecular and cortical bones.\cite{BOSSY04,LEVEQUE12} The anisotropy of a medium changes only the $4\times 4$ poroelastic tensor $\bm{C}$ (\ref{prop:nrj}). So the ADER 4 scheme and the immersed interface method have to be modified, but the approximation of the fractionnal derivative is still valid.
\item \emph{Thermic boundary-layer.} In cases where the saturating fluid is a gas, thermo-mechanical effects have to be taken into account. Extended versions of the Biot-JKD have been developed,\cite{LAFARGE97} involving additional order 1/2 fractional derivatives. The numerical method developed in this paper should lend itself well to working with this model.
\end{itemize}

%------------------------------------------------------------------------------------------
%------------------------------------------------------------------------------------------

\appendix

\section{Matrices of propagation and dissipation}\label{annexe:matABS}

The matrices in (\ref{eq:syst_hyperbolique}) are
\begin{equation}
\begin{array}{l}
{\bf A} = \left( 
\begin{array}{c|c|c}
\bf{0}_{4,4} & \bf{A}_1 & \bf{0}_{4,2N}\\ 
[5pt]\hline
\bf{A}_2 & \bf{0}_{4,4} & \bf{0}_{4,2N}\\
[5pt]\hline
\bf{0}_{2N,4} & \bf{A}_3 & \bf{0}_{2N,2N}
\end{array}
\right) ,\quad
{\bf A}_1 = \left( 
\begin{array}{cccc}
\displaystyle -\frac{\rho_w}{\chi} & 0 & 0 & \displaystyle -\frac{\rho_f}{\chi}\\
0 & \displaystyle -\frac{\rho_w}{\chi} & 0 & 0\\
\displaystyle \frac{\rho_f}{\chi} & 0 & 0 & \displaystyle \frac{\rho}{\chi}\\
0 & \displaystyle \frac{\rho_f}{\chi} & 0 & 0
\end{array}
\right) ,\\
[50pt]
{\bf A}_2 = \left( 
\begin{array}{cccc}
\displaystyle -(\lambda_f+2\,\mu) & 0 & \displaystyle -m\,\beta & 0\\
0 & \displaystyle -\mu & 0 & 0\\
\displaystyle -\lambda_f & 0 & \displaystyle -m\,\beta & 0\\
\displaystyle m\,\beta & 0 & m & 0
\end{array}
\right) ,\quad {\bf A}_3 = \left( 
\begin{array}{cccc}
\displaystyle \frac{\rho_f}{\chi} & 0 & 0 & \displaystyle \frac{\rho}{\chi}\\
0 & \displaystyle \frac{\rho_f}{\chi} & 0 & 0\\
\vdots & \vdots & \vdots & \vdots \\
\displaystyle \frac{\rho_f}{\chi} & 0 & 0 & \displaystyle \frac{\rho}{\chi}\\
0 & \displaystyle \frac{\rho_f}{\chi} & 0 & 0
\end{array}
\right);
\end{array}
\label{eq:matriceA}
\end{equation}

\begin{equation}
\begin{array}{l}
{\bf B} = \left( 
\begin{array}{c|c|c}
\bf{0}_{4,4} & \bf{B}_1 & \bf{0}_{4,2N}\\ 
[5pt]\hline
\bf{B}_2 & \bf{0}_{4,4} & \bf{0}_{4,2N}\\
[5pt]\hline
\bf{0}_{2N,4} & \bf{B}_3 & \bf{0}_{2N,2N}
\end{array}
\right) ,\quad
{\bf B}_1 = \left( 
\begin{array}{cccc}
0 & \displaystyle -\frac{\rho_w}{\chi} & 0 & 0 \\
0 & 0 & \displaystyle -\frac{\rho_w}{\chi} & \displaystyle -\frac{\rho_f}{\chi}\\
0 & \displaystyle \frac{\rho_f}{\chi} & 0 & 0\\
0 & 0 & \displaystyle \frac{\rho_f}{\chi} & \displaystyle \frac{\rho}{\chi}
\end{array}
\right) ,\\
[50pt]
{\bf B}_2 = \left( 
\begin{array}{cccc}
0 & \displaystyle -\lambda_f & 0 & \displaystyle -m\,\beta\\
\displaystyle -\mu & 0 & 0 & 0\\
0 & \displaystyle -(\lambda_f+2\,\mu) & 0 & \displaystyle -m\,\beta\\
0 & \displaystyle m\,\beta & 0 & \displaystyle m
\end{array}
\right) ,\quad {\bf B}_3 = \left( 
\begin{array}{cccc}
0 & \displaystyle \frac{\rho_f}{\chi} & 0 & 0\\
0 & 0 & \displaystyle \frac{\rho_f}{\chi} & \displaystyle \frac{\rho}{\chi}\\
\vdots & \vdots & \vdots & \vdots \\
0 & \displaystyle \frac{\rho_f}{\chi} & 0 & 0\\
0 & 0 & \displaystyle \frac{\rho_f}{\chi} & \displaystyle \frac{\rho}{\chi}
\end{array}
\right);
\end{array}
\label{eq:matriceB}
\end{equation}
\begin{equation}
\begin{array}{l}
{\bf S} = \left( \begin{array}{c|c|c}
{\bf 0}_{4,4} & {\bf 0}_{4,4} & {\bf S}_1\\
[5pt]\hline
{\bf 0}_{4,4} & {\bf 0}_{4,4} & {\bf 0}_{4,2N}\\
[5pt]\hline
{\bf S}_3 & {\bf 0}_{2N,4} & {\bf S}_2
\end{array}
\right),\quad
{\bf S}_3 = \left( 
\begin{array}{cccc}
0 & 0 & -\Omega & 0\\
0 & 0 & 0 & -\Omega \\
\vdots & \vdots & \vdots & \vdots \\
0 & 0 & -\Omega & 0\\
0 & 0 & 0 & -\Omega
\end{array}
\right) ,\\
[50pt]
{\bf S}_1 = \left( 
\begin{array}{ccccc}
\displaystyle -\frac{\rho_f}{\rho}\,\gamma\,a_1 & 0 & ... & \displaystyle -\frac{\rho_f}{\rho}\,\gamma\,a_N & 0 \\
[5pt]
0 & \displaystyle -\frac{\rho_f}{\rho}\,\gamma\,a_1 & ... & 0 & \displaystyle -\frac{\rho_f}{\rho}\,\gamma\,a_N \\
[5pt]
\displaystyle \gamma\,a_1 & 0 & ... & \displaystyle \gamma\,a_N & 0 \\
[5pt]
0 & \displaystyle \gamma\,a_1 & ... & 0 & \displaystyle \gamma\,a_N
\end{array}
\right),\\
[50pt]
{\bf S}_2 = \left( 
\begin{array}{ccccc}
\gamma\,a_1 + (\theta_1+\Omega) & 0 & ... & \gamma\,a_N & 0 \\
[5pt]
0 & \gamma\,a_1 + (\theta_1+\Omega) & ... & 0 & \gamma\,a_N \\
[5pt]
 \vdots & \vdots & \vdots & \vdots & \vdots \\
[5pt]
\gamma\,a_1 & 0 & ... & \gamma\,a_N + (\theta_N+\Omega) & 0 \\
[5pt]
0 & \gamma\,a_1 & ... & 0 & \gamma\,a_N + (\theta_N+\Omega)
\end{array}
\right).
\end{array}
\label{eq:matriceS}
\end{equation}

%------------------------------------------------------------------------------------------
%------------------------------------------------------------------------------------------

%\bibliography{mybiblio}

\end{document}